\documentclass[12pt,fleqn]{article}

\usepackage{graphicx}
\usepackage{pdftricks}
\usepackage[utf8]{inputenc}
\usepackage{longtable}
\RequirePackage{lineno} 
\usepackage{epstopdf}

\begin{document}

\title{\Large Solid-like features in dense vapors near the fluid critical point}

\author{George Ruppeiner,\footnote{Division of Natural Sciences, New College of Florida, 5800 Bay Shore Road, Sarasota, FL 34243, USA (ruppeiner@ncf.edu)} $\,$ Nathan Dyjack, Abigail McAloon, and Jerry Stoops}

\maketitle

\begin{abstract}
The phase diagram (pressure versus temperature) of the pure fluid is typically envisioned as being featureless apart from the presence of the liquid-vapor coexistence curve terminating at the critical point.  However, a number of recent authors have proposed that this simple picture misses important features, such as the Widom line, the Fisher-Widom line, and the Frenkel line.  In our paper we discuss another way of augmenting the pure fluid phase diagram, lines of zero thermodynamic curvature $R=0$ separating regimes of fluid solid-like behavior ($R>0$) from gas-like or liquid-like behavior ($R<0$).  We systematically evaluate $R$ for the $121$ pure fluids in the NIST/REFPROP (version 9.1) fluid database near the saturated vapor line from the triple point to the critical point.  Our specific goal was to identify regions of positive $R$ abutting the saturated vapor line (''feature D'').  We found: a) $97/121$ of the NIST/REFPROP fluids have feature D.  b) The presence and character of feature D correlates with molecular complexity, taken to be the number of atoms $Q$ per molecule.  c) The solid-like properties of feature D might be attributable to a mesoscopic model based on correlations among coordinated spinning molecules, a model that might be testable with computer simulations.  d) There are a number of correlations between thermodynamic quantities, including the acentric factor $\omega$, but we found little explicit correlation between $\omega$ and the shape of a molecule.  e) Feature D seriously constrains the size of the asymptotic fluid critical point regime, possibly resolving a long-standing mystery about why these are so small.  f) Feature D correlates roughly with regimes of anomalous sound propagation.
\end{abstract}

\\

\section{Introduction}

\par
Fluid thermodynamic properties are readily measurable, and large data sets have been collected.  Much of this data has been fit to convenient functional forms, typically for the molar Helmholtz free energy, and made available on comprehensive web sites.  In our paper, we use the NIST Chemistry WebBook \cite{NIST, Lemmon2014} to compute the thermodynamic curvature $R$ for a broad range of fluids in the general vicinity of the fluid critical point, though not in the asymptotic critical region.  Of specific interest are anomalous regions of positive $R$ in the dense vapor phase that we attribute to the formation of solid-like repulsive clusters.

\par
Several previous studies of $R$ were done along the liquid-vapor coexistence curve, and its extension into the supercritical region (the Widom line) \cite{Ruppeiner2012b, Ruppeiner2012a, May2012}.  More generally, $R$ was calculated for the full fluid phase diagram for four fluids \cite{Ruppeiner2015a}, and for thermodynamic data collected from computer simulations of the Lennard-Jones fluid \cite{May2013}.  There was also an investigation of $R$ in water near a conjectured second critical point in the metastable liquid phase \cite{May2015}.  The picture that emerged from these efforts is consistent and compelling.  In our paper we add to this picture results obtained by looking at $121$ pure fluids in the dense vapor phase.

\par
We are certainly aware that fits to fluid data are not free from uncertainty since the filling of fluid regimes with data is incomplete and uneven.  There are also regimes near the critical point that pose special experimental and theoretical difficulties.  In addition, fit quality is generally tested against data only up to second-order derivatives of the Helmholtz free energy, short of the third-order derivatives present in the curvature $R$.  Hence, specific results for any of our fluids might be challenged.

\par
However, the large fluid data fitting project underlying NIST/REFPROP represents results likely to be stable against major near term changes.  We thus view our present project as timely, and convey the hope that our ''big data'' effort over the totality of NIST/REFPROP, with no pause to examine the detailed goodness of fit for any particular fluid, transmits ideas that are essentially correct, and that point the way to further productive research.

\par
Fluid properties result from the interactions among many molecules.  Basic principles of intermolecular interactions is a topic that has seen considerable recent advances, particularly via quantum computational methods.  But, the transition from the microscopic regime to the macroscopic thermodynamic regime poses great difficulties.  Beyond rarefield gases, statistical mechanics \cite{Landau1980, Pathria2011} offers only approximation schemes, and the solution to microscopic models specifically devised for ready solution.  Of course, computer simulations provide powerful mimics of real systems, but even these can overlook fundamental insight.

\par
Our paper further develops a relatively new approach to fluids, where instead of building up from the microscopic level, as in statistical mechanics, we build down from the macroscopic level to the mesoscopic level.  The idea is to relate important mesoscopic patterns to readily measured thermodynamic properties.  It has been argued that thermodynamic fluctuation theory, augmented by the logically necessary thermodynamic curvature $R$, offers the necessary link between the macroscopic and the mesoscopic.

\par
The thermodynamic curvature $R$, always calculated at the macroscopic level, measures the size scale of mesoscopic fluctuations.  Very near the critical point, $|R|$ gives the diverging correlation length $\xi$: $|R|\propto\xi^3$ \cite{Ruppeiner1995, Ruppeiner1979, Johnston2003}.  Elsewhere, even at length scales as small as cubic nanometers, $R$ gives the approximate volume of organized mesoscale structures.  The sign of $R$ is also significant: $R$ is positive/negative for states where repulsive/attractive intermolecular interactions dominate.

\par
Most of a fluid's temperature-pressure phase diagram corresponds to average molecular separation distances sufficiently large that the attractive part of the intermolecular interaction potential dominates.  Thus, $R$ in fluids is found to be mostly negative, diverging to negative infinity at the critical point.  Fluid states with positive $R$ are somewhat special, and typically occur at large densities, in solid-like states dominated by repulsive intermolecular interactions.

\par
Of interest in our paper, however, are instances of positive fluid $R$ in the general vicinity of the critical point.  We found such features in the majority of the $121$ fluids we examined.  Our paper is a comprehensive look at phenomena spanning a range of fluids.  Another recent large scale fluid tabulation has been the determination of the Gr{\"u}neisen parameter for $28$ fluids \cite{Mausbach2016}.

\section{Theory and background}

In this section we give a theoretical background.

\subsection{Thermodynamic geometry}

Thermodynamics is typically defined in the thermodynamic limit in which we have a uniform system characterized by a few macroscopic parameters \cite{Callen1985}.  For the pure fluid, we have the internal energy $U$, the number of particles $N$, the volume $V$, and the entropy $S$.  The fundamental thermodynamic equation sets $U=U(S, N, V)$.  Define the temperature $T=U_{,S}$, the pressure $p=-U_{,V}$, and the chemical potential $\mu=U_{,N}$, where the comma notation indicates partial differentiation.

\par
Define the Helmholtz free energy $A = U - T S$, with this quantity per volume $f=A/V$.  We may naturally write $f=f(T,\rho)$, an expression which yields all of the thermodynamic properties \cite{Callen1985}.  Here, $\rho=N/V$ is the particle density.  We also have the entropy per volume $s=S/V=-f_{,T}$, and the chemical potential $\mu=f_{,\rho}$.  The energy per volume $u = U/V = f + T s$, and $p=-u + T s + \mu\rho$.

\par
Consider some open subsystem $\mathcal{A}$, with fixed volume $V$, of a very large environment $\mathcal{A}_0$.  $\mathcal{A}_0$ has a fixed thermodynamic state with temperature and density $(T_0,\rho_0)$.  The thermodynamic state $(T,\rho)$ of $\mathcal{A}$ fluctuates as particles and energy are randomly exchanged with its environment.  As was first argued by Einstein in 1904, the probability density for finding the state of $\mathcal{A}$ in the small range $(T,\rho)$ to $(T+dT,\rho+d\rho)$ is proportional to \cite{Landau1980,Pathria2011}

\begin{equation}P\,dT\,d\rho\propto\exp\left(S_{total}/k_B\right)dT\,d\rho,\label{10}\end{equation}

\noindent where $S_{total}$ is the entropy of $\mathcal{A}_0$ when $\mathcal{A}$ is in the state with coordinates $(T,\rho)$, and $k_B$ is Boltzmann's constant.

\par
To write the probability density in Eq. (\ref{10}) in familiar thermodynamic terms, expand $S_{total}$ to second order around its maximum corresponding to $\{T,\rho\}=\{T_0,\rho_0\}$.  Standard methods \cite{Landau1980, Pathria2011, Ruppeiner1995} yield:

\begin{equation}P\propto\mbox{exp}\left(-\frac{V}{2}\Delta\ell ^2\right),\label{20}\end{equation}

\noindent where the thermodynamic metric

\begin{equation}\Delta\ell^2=\frac{1}{k_B T}\left(\frac{\partial s}{\partial T}\right)_{\rho}\Delta T^2+\frac{1}{k_B T}\left(\frac{\partial\mu}{\partial\rho}\right)_T\Delta\rho^2,\label{30}\end{equation}

\noindent with $\Delta T=T-T_0$, and $\Delta\rho=\rho-\rho_0$.  The coefficients of $\Delta T^2$ and $\Delta\rho^2$ in Eq. (\ref{30}) may be evaluated in either of the states $(T_0,\rho_0)$ or $(T,\rho)$, since small fluctuations always have these states close together.  Below, we evaluate these metric coefficients at $(T,\rho)$.

\par
The thermodynamic metric Eq. (\ref{30}) induces a thermodynamic curvature $R$ on the manifold of thermodynamic states \cite{Ruppeiner1995,Ruppeiner1979}:

\begin{equation}R=\frac{1}{\sqrt{g}}\left[\frac{\partial}{\partial T}\left(\frac{1}{\sqrt{g}}\frac{\partial g_{\rho\rho}}{\partial T}\right) + \frac{\partial}{\partial \rho}\left(\frac{1}{\sqrt{g}}\frac{\partial g_{TT}}{\partial \rho}\right)\right],\label{40}\end{equation}

\noindent where

\begin{equation}g_{TT}=\frac{1}{k_B T}\left(\frac{\partial s}{\partial T}\right)_{\rho},\label{50}\end{equation}

\begin{equation}g_{\rho\rho}=\frac{1}{k_B T}\left(\frac{\partial\mu}{\partial\rho}\right)_T,\label{60}\end{equation}

\noindent and

\begin{equation} g=g_{TT}\,g_{\rho\rho}.\label{70}\end{equation}

\par
There is substantial evidence \cite{Ruppeiner2012b, Ruppeiner2012a, May2012, Ruppeiner2015a, May2013, May2015, Ruppeiner1995, Ruppeiner1979, Johnston2003, Ruppeiner2010} that $R$ is a thermodynamic measure of intermolecular interactions.  It appears that $|R|$ gives the characteristic size of organized mesoscopic fluctuations in fluid and magnetic systems.  In the asymptotic critical region, this size is given by the correlation length $\xi$:

\begin{equation}\xi^3=-\frac{1}{2}\,R.\label{80}\end{equation}

\par
The sign of $R$ also appears to be significant: $R$ is positive/negative for systems in states dominated by repulsive/attractive intermolecular interactions.  $R$ for systems with no interactions between microscopic constituents, such as the pure ideal gas and the paramagnet, have $R=0$.  These general findings have been verified in a number of contexts.\footnote{Since the analysis of ref. \cite{Ruppeiner1979}, the values of the critical parameters for the correlation length have been upgraded for seven simple fluids \cite{Garrabos2006}, a study that might lead to another check of Eq. (\ref{80}).}

\par
Metric geometry in thermodynamics originated with Weinhold \cite{Weinhold1975}, and was applied as well to finite-time thermodynamics; see \cite{Andresen2015} for a brief recent review.  For a recent discussion of the use of thermodynamic curvature in quantum statistical mechanics; see \cite{Ubriaco2016}.

\subsection{Extended features of the fluid state}

A theme in our paper is the identification of curves separating regimes of attractive and repulsive intermolecular interactions.  An early attempt to do this consists of the Fisher-Widom line marking the boundary between regimes of monotonic/oscillatory long-range decay of the pair correlation function $G(r)$, corresponding to attractive/repulsive intermolecular interactions \cite{Fisher1969}.  This correspondence was demonstrated explicitly by calculations on several continuum models.

\par
\begin{figure}
\centering
\includegraphics[width=9cm]{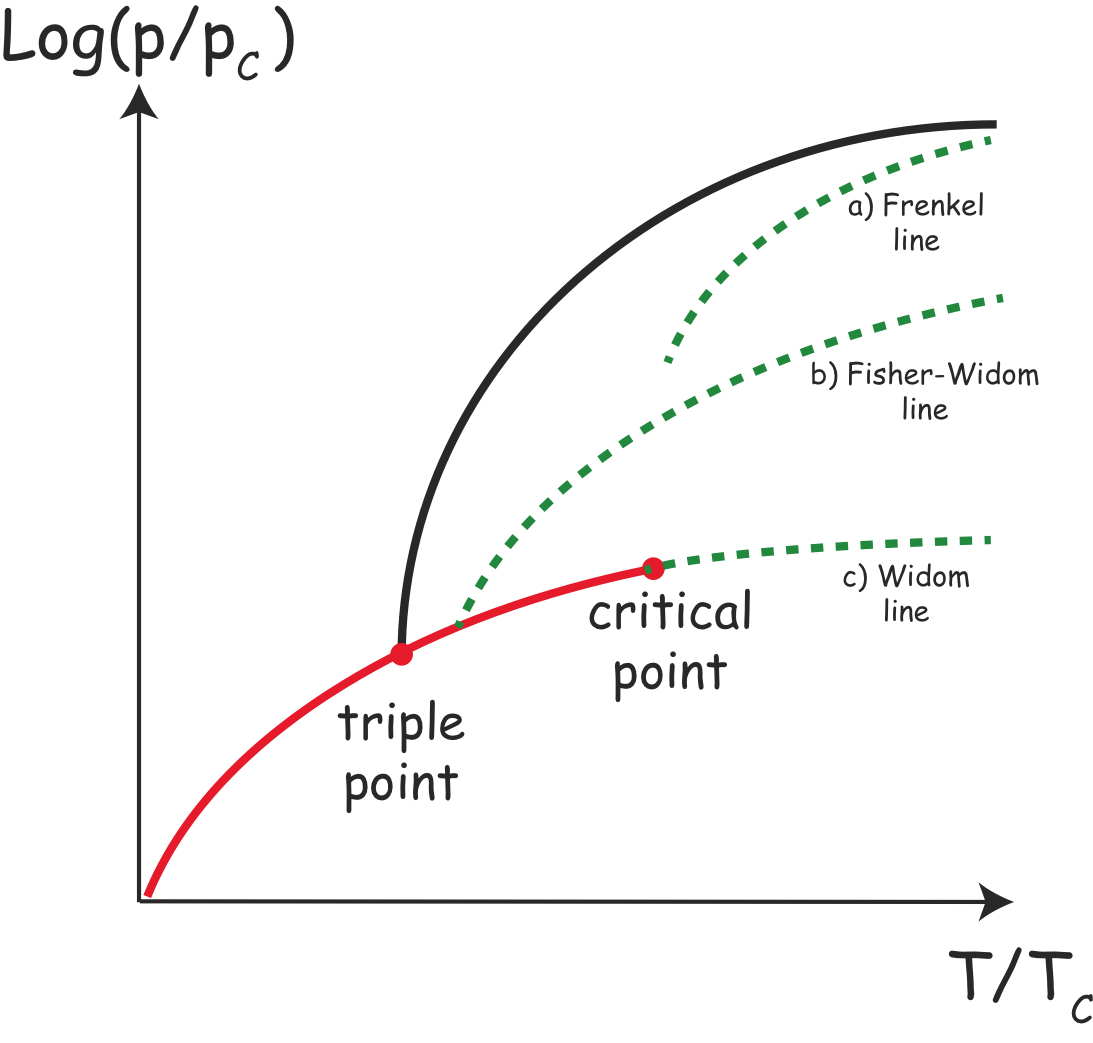}
\caption{Schematic diagrams of three curves supplementing the fluid phase diagram: a) the conjectured Frenkel line, defined by the changing character of solid-like oscillations; b) the Fisher-Widom line representing the crossover in the long-range pair correlation function decay from oscillatory to monotonic; and c) the Widom line representing the smooth continuation of the boiling curve into the supercritical regime.}
\label{fig:1}
\end{figure}

Evans {\it et al.} \cite{Evans1993} expanded on the Fisher-Widom argument, and Ruppeiner and Chance \cite{Ruppeiner1990} discussed the Fisher-Widom line for the 1D Takahashi gas in the context of $R$.  But a difficulty implementing this agenda with model calculations is that it requires a calculation of $G(r)$, and this is difficult to do beyond a few models in one dimension.  Perhaps modern computer simulation methods, for which a number of methods exist for calculating pair correlation functions, even for larger molecules, offer an update for this method.  The Fisher-Widom line is shown schematically in Figure \ref{fig:1}.

\par
Another fluid line receiving recent attention is the Widom line \cite{McMillan2010, Simeoni2010, Brazhkin2011, Brazhkin2014, Corradini2015}, representing the smooth continuation of the liquid-vapor coexistence curve into the supercritical region.  The Widom line is defined as the curve along which $\xi$ has a maximum.  Experimental evidence for an abrupt crossover of certain dynamical fluid properties from gas-like on the low-pressure side to liquid-like on the high-pressure side has been reported along the Widom line \cite{Simeoni2010}.

\par
Since it is difficult to compute $\xi$, the Widom line is usually determined using maxima in static thermodynamic response functions, such as the heat capacity.  However, it was argued that, because of the proportionality Eq. (\ref{80}), peaks in the readily calculated $R$ offer a better measure of the Widom line than response functions \cite{Ruppeiner2012b, May2012}.  The Widom line is shown schematically in Figure \ref{fig:1}.

\par
Yet another proposed addition to the fluid phase diagram is the Frenkel line, which is defined by the disappearance of solid-like oscillations in particle dynamics.  The Frenkel line marks a dynamic transition between gas-like and liquid-like behavior, for which experimental evidence has been reported \cite{Bolmatov2013, Bolmatov2015}.  The Frenkel line is shown schematically in Figure \ref{fig:1}.

\par
Also yielding special curves in the dense fluid and solid states are Roskilde-simple systems, in which the simplicity of the repulsive intermolecular interaction yields universal curves when expressed in terms of appropriately reduced thermodynamic variables \cite{Dyre2014}.  This idea applies as well to real systems.

\subsection{Mesoscopic structures in the fluid state}

\par
\begin{figure}
\centering
\includegraphics[width=9cm]{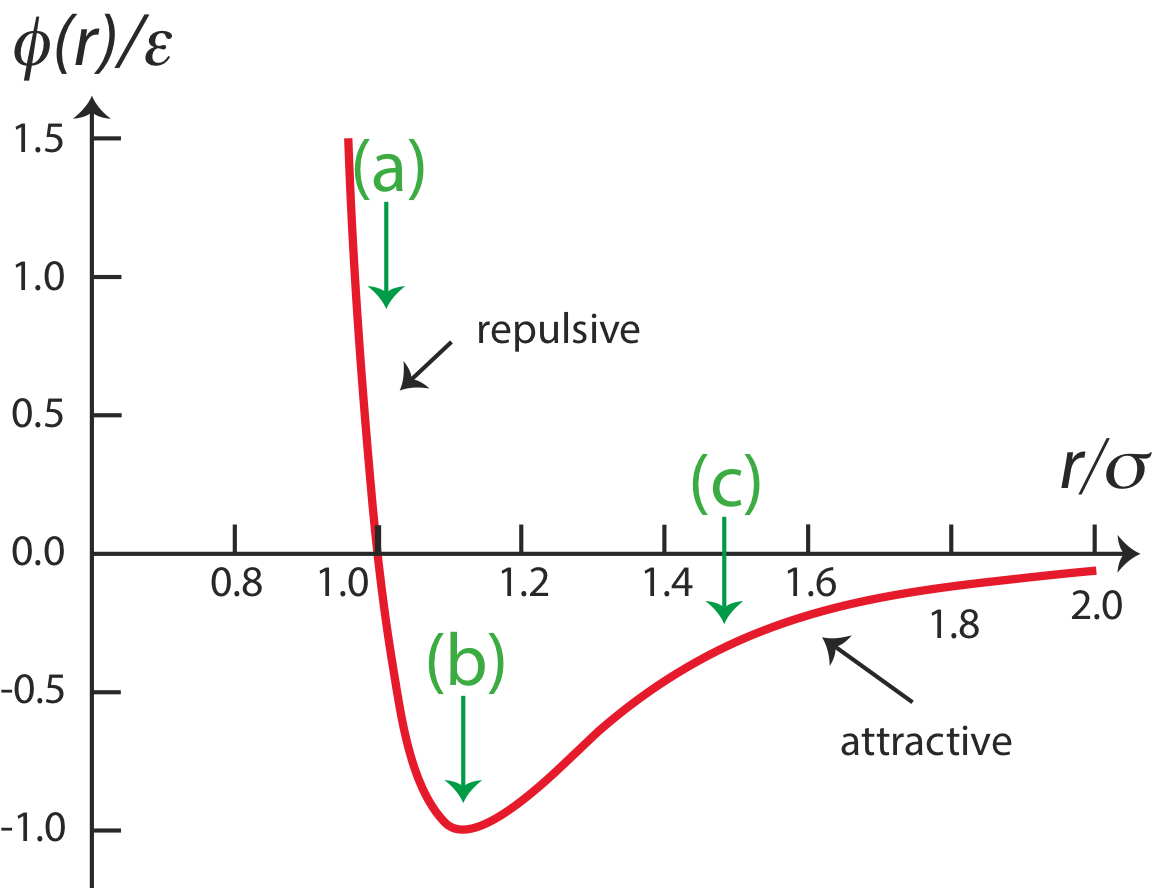}
\caption{The Lennard Jones pair potential $\phi(r)$ between two particles in a three-dimensional fluid.  The down-pointing arrows indicate: (a) $r/\sigma = 1.01$, where computer simulation results indicate the approximate location of the supercritical $R=0$ curve for a range of reasonable $T$'s; (b) $r/\sigma=1.12246$ where $\phi(r)$ has its minimum; and (c) $r/\sigma=1.48$, where computer simulations place the critical density.}
\label{fig:2}
\end{figure}

Useful for further discussions of fluids is a computer fluid composed of particles interacting via the Lennard-Jones (LJ) pair potential:

\begin{equation}\phi(r)=4\epsilon\left[\left(\frac{\sigma}{r}\right)^{12}-\left(\frac{\sigma}{r}\right)^6\right],\label{90}\end{equation}

\noindent where $r$ is the distance between two particles, and $\sigma$ and $\epsilon$ are size and energy parameters.  Figure \ref{fig:2} shows this LJ pair potential, which offers a good approximation of the potential in real fluids.  The down pointing arrows denote: a) approximately where LJ computer simulations indicate $R=0$ in the supercritical regime for a range of reasonable $T$'s \cite{May2013}; b) where the pair potential has its minimum; and c) where LJ computer simulations indicate the critical density \cite{Thol2016}.

\par
For real fluids in the gaseous state, and with large intermolecular separation distance $r$, the attractive part of the intermolecular potential dominates, and $R$ is negative.  Attractive interactions also dominate near the critical point, at which $R\to -\infty$.  In the condensed liquid or solid states, where the molecules are essentially touching each other, we typically find $|R|^{1/3}$ on the order of a molecular separation distance.  As in the LJ fluid, the corresponding sign of $R$ may be positive or negative, ''solid-like'' or ''liquid-like,'' respectively.  The former case typically has the larger densities, and has the molecules in the regime of repulsive interactions.

\par
\begin{figure}
\centering
\includegraphics[width=9cm]{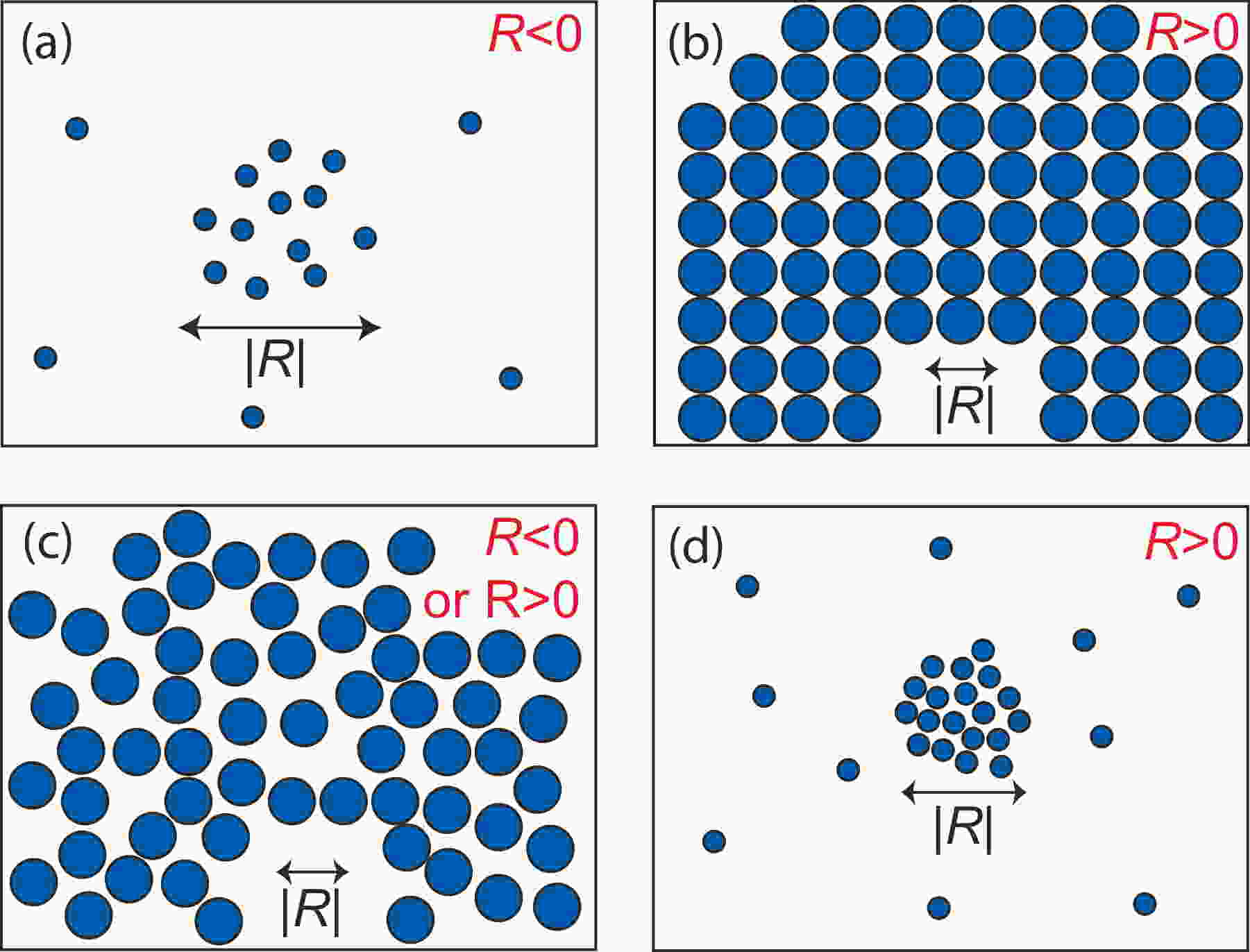}
\caption{Schematics of four organized mesoscopic particle configurations and their $R$'s: (a) an attractive near critical point cluster, with $R<0$ and $|R|\sim\xi^3$; (b) the solid state, with positive $R\sim v$, and with $v$ the molecular volume; (c) the compact liquid state, with positive or negative $R$ and $|R|\sim v$; and (d) a repulsive solid-like cluster held up by close-range particle repulsion, with positive $R\sim$ cluster volume.}
\label{fig:3}
\end{figure}

More difficult to interpret physically are anomalous states of positive $R$ in the vicinity of the critical point, with densities less than the condensed densities.  We interpret such solid-like states in terms of fluctuating ''repulsive clusters,'' shown schematically in Figure \ref{fig:3}, along with three other characteristic mesoscopic structures.  Identifying repulsive cluster states in a broad range of dense vapors is the main point of our paper.  One of us proposed \cite{Ruppeiner2012a} that the average repulsive cluster volume is given by $R$, but the only previous case reported of positive $R$ in the dense vapor phase was in water \cite{Ruppeiner2015a}, with clusters identified in computer simulations \cite{Johansson2005}.

\par
\begin{figure}
\centering
\includegraphics[width=9cm]{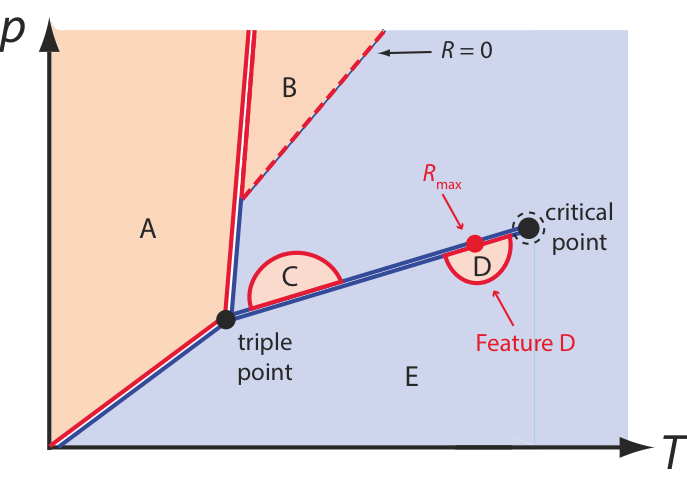}
\caption{A schematic fluid phase diagram, showing the triple point, the critical point, the sublimation curve, the melting curve, and the boiling curve.  The red/blue coloring indicates whether the sign of $R$ is positive/negative.  Feature A is the solid phase, features B and C are in the liquid phase, but with conjectured solid-like properties, feature D corresponds to the repulsive cluster phase of primary interest in our paper, and feature E is the majority of the gas phase.  The red dot $R_{max}$ denotes the location of the maximum value of $R$ within feature D.  The dotted circle around the critical point indicates the asymptotic critical regime, contracted significantly by the presence of feature D.  Boundaries between features with opposite $R$ signs not on a coexistence curve do not have sharp thermodynamic discontinuities.}
\label{fig:4}
\end{figure}

To form repulsive clusters requires a reasonably high density, else the molecular collision rate is too low to overcome the natural tendency for these clusters to break up.  But we do not want the density so high as to be in the liquid or solid state, which are quite different.  Our suggestion here is that conditions for repulsive clusters might hold in the dense vapor phase, as shown schematically in Fig. \ref{fig:4}, where this state is referred to as feature D.

\par
To summarize, the repulsive cluster state differs substantially from the solid state, and from the solid-like liquid state.  The latter two states have larger densities, and positive $R$ values on the order of the volume of a molecule.  In contrast, the repulsive cluster state corresponds to lower densities, characteristic more of a dense gas, and has $R$ which may become much larger than the volume of a molecule.

\subsection{Feature D and the BZT region}
Feature D might relate to sound propagation anomalies in dense vapors near the liquid-vapor coexistence curve.  The fluid sound velocity $c$ is given by the isentropic compressibility \cite{Landau1975}:

\begin{equation}c^2=-\nu^2\left(\frac{\partial p}{\partial \nu}\right)_s, \label{100}\end{equation}

\noindent where $\nu$ and $s$ denote the volume and entropy per mass, respectively.  Anomalous sound propagation results from the dependence of $c$ on $\nu$, resulting in variable sound speed along the wave train.  The result is a tendency to form shock waves, a tendency measured by the fundamental derivative of gas dynamics \cite{Colonna2006}:

\begin{equation}\Gamma=1-\frac{\nu}{c}\left(\frac{\partial c}{\partial \nu}\right)_s\label{110}.\end{equation}

\noindent For $\Gamma<0$, there is the possibility of ''nonclassical'' rarefield shock waves in regions referred to as Bethe, Zel'dovich, and Thompson (BZT) regions, originally identified in the van der Waals model (vdW) in dense saturated vapors.

\par
Colonna and Guardone \cite{Colonna2006} correlated the BZT region in vdW with increasing molecular complexity, and with increasing constant volume molar specific heat capacity $c_{\nu}$.  Their measure of molecular complexity was the number of classical degrees of freedom (translational, vibrational, and rotational):

\begin{equation}N^{avail}=6\,Q-5-\epsilon,\label{120}\end{equation}

\noindent where $Q$ is the number of atoms per molecule, and $\epsilon=-2,0,1$ for monatomic, linear polyatomic, and nonlinear polyatomic molecules, respectively.  $N^{avail}$ only sets an upper limit for the number of degrees of freedom, which might not all be active at low temperatures.  In our study, it turns out that $Q$ is our best measure of complexity, so we essentially agree with the choice of complexity measure in \cite{Colonna2006}, since the $Q$ term dominates in $N^{avail}$.

\par
The BZT study was later extended to real gasses having parametrized fit equations of state \cite{Colonna2009}, and to a database of $1,800$ pure fluids represented by simple models \cite{Castier2012}.  The later study found 185 organic molecules having BZT regions, with most of the BZT molecules having $10$ or more carbon atoms, more carbon atoms than most of the molecules considered here.

\par
In terms of rough location and dependence on molecular complexity, the BZT region correlates at least qualitatively with feature D.  An additional similarity is that $\Gamma$ and $R$ both contain up to three derivatives of the Helmholtz free energy.  Perhaps, feature D and the BZT region have some common mesoscopic mechanism.

\subsection{Measures of molecular complexity}

\par
The character of feature D correlates with molecular complexity, for which we considered three measures.  The first is the molecular weight $MW$.  For organic molecules, $MW$ correlates roughly with molecular complexity, since, typically, a larger $MW$ means more atoms per molecule.  But $MW$ could be misleading, since a simple molecule composed of heavy atoms might have a large $MW$, e.g., monatomic krypton has $MW=83.80$ g/mol, and polyatomic cyclohexane has $MW=84.16$ g/mol.

\par
Our second complexity measure is the weight-independent number of atoms $Q$ per molecule, determined by consulting PubChem \cite{Kim2015}.  With $Q$ reasonably large, $Q$ is essentially a measure equivalent to $N^{avail}$ in Eq. (\ref{120}).

\par
Our third complexity measure is the Wiener index $Wi$, which in chemical graph theory is a topological index of a molecule \cite{Wiener1947, Mohar1988}.  $Wi$ is defined as the sum of the lengths of the shortest paths between all pairs of connected vertices in the chemical graph of the molecule.\footnote{A pair of adjacent atoms in a molecule contribute $1$ to the path length if they are connected by a bond, and $0$ otherwise.}  Sometimes the hydrogen atoms are left out of this measure, but since they are included in both $MW$ and $Q$, we will include them as well in our $Wi$ calculations.

\par
Another microscopic measure of a molecule is its shape, which we classified according to its spatial molecular dimension $d$: 0, 1, 2, or 3.  All our molecules were visually inspected in PubChem \cite{{Kim2015}}, and assigned a value of $d$.  In some cases, $d$ was clear; for example, argon ($d=0$), decane ($d=1$), benzene ($d=2$), and isobutane ($d=3$).  Spatial dimension was estimated visually for molecules with complex structural arrangements.

\par
An attempt at a thermodynamic measure of $d$ is the acentric factor $\omega$ defined as

\begin{equation}\omega =-\log_{10}\left(\frac{p^{\rm{sat}}}{p_c}\right)-1, \label{130} \end{equation}

\noindent where $p^{\rm{sat}}$ is the saturated vapor pressure evaluated at $T/T_c=0.7$ \cite{Valderrama2007}, and $p_c$ is the critical pressure.  $\omega$ is intended as a measure of the non-sphericity of molecules, and it is defined so that monotonic atoms ($d=0$) have $\omega$ near zero.

\subsection{Corresponding states $\mathcal{P}_{max}$ of maximum $R$}

Identifying and characterizing the fluids with feature D is the major theme of our paper.  But we may also make systematic comparisons between Dfluids (fluids with feature D), provided we identify corresponding reference states between the Dfluids.  For each Dfluid, we picked the point $\mathcal{P}_{max}$ having the maximum value $R_{max}$ of $R$ in feature D.  $\mathcal{P}_{max}$ always seemed to lie on the saturated vapor curve itself, and all of our reported $R_{max}$ values were located there.

\par
We might naively expect some form of common behavior between Dfluids at $\mathcal{P}_{max}$, even though it might be hard to specify what such behavior is without specific knowledge of the mesoscopic mechanism behind feature D.  For selecting a reference state, an advantage of $R$ over other thermodynamic functions is not just its direct connection to mesoscopic fluctuations, but the existence of a local maximum for $R$.

\subsection{The alkanes, perfluoroalkanes, and repulsive clusters}

Although $R_{max}$ allows for overall comparisons between all the $97$ Dfluids, emerging relationships could be somewhat difficult to interpret physically because of substantial differences among the Dfluid molecules.  Families of related molecules allow more focused comparisons.  Our primary example consists of the linear alkanes.

\par
\begin{figure}
\centering
\includegraphics[width=7cm]{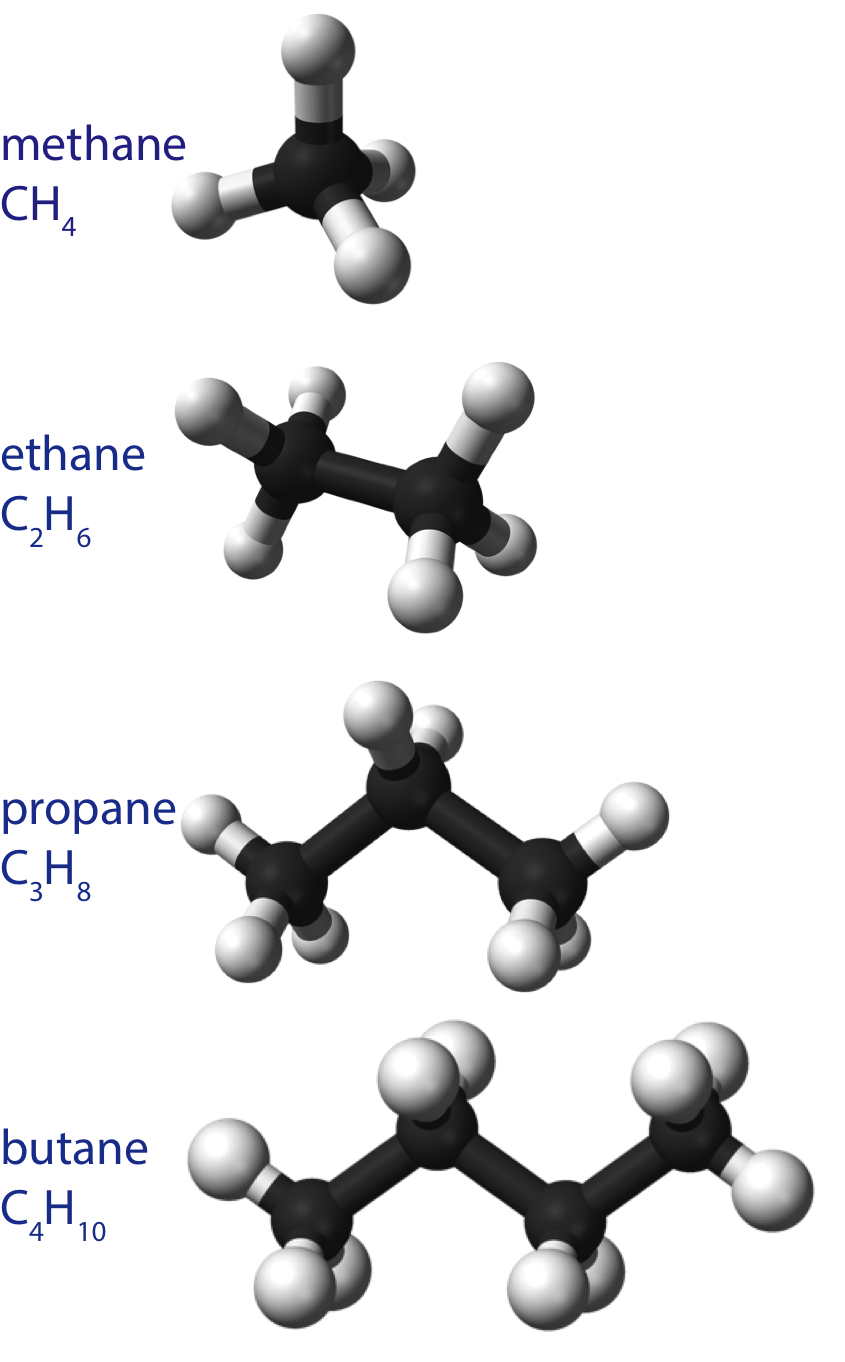}
\caption{The first four members of the linear alkane series of molecules, with chemical formulas $C_n H_{2n+2}$ ($n=1,2,3,\ldots$).  Except for methane, all the linear alkanes are Dfluids.}
\label{fig:5}
\end{figure}

The linear alkanes start with methane CH$_4$, and can be thought of as extended progressively by adding a carbon atom and two hydrogen atoms at a time; see Figure \ref{fig:5}.  A similar sequence consists of the perfluoroalkanes, with fluorine atoms in place of the hydrogen atoms in the alkanes.  The alkane Dfluids that we looked at were ethane \cite{Buecker2006a}, propane \cite{Lemmon2009}, butane \cite{Buecker2006b}, pentane \cite{Span2003}, hexane \cite{Span2003}, heptane \cite{Span2003}, octane \cite{Span2003}, nonane \cite{Lemmon2006}, decane \cite{Lemmon2006}, undecane \cite{Aleksandrov2011}, and dodecane \cite{Lemmon2004}.  The perfluoroalkane Dfluids that we looked at were hexafluoroethane (R116) \cite{Lemmon2006}, octafluoropropane (R218) \cite{Lemmon2006}, perfluorobutane \cite{Huber1994}, and perfluoropentane \cite{Huber1994}.

\par
\begin{figure}
\centering
\includegraphics[width=7cm]{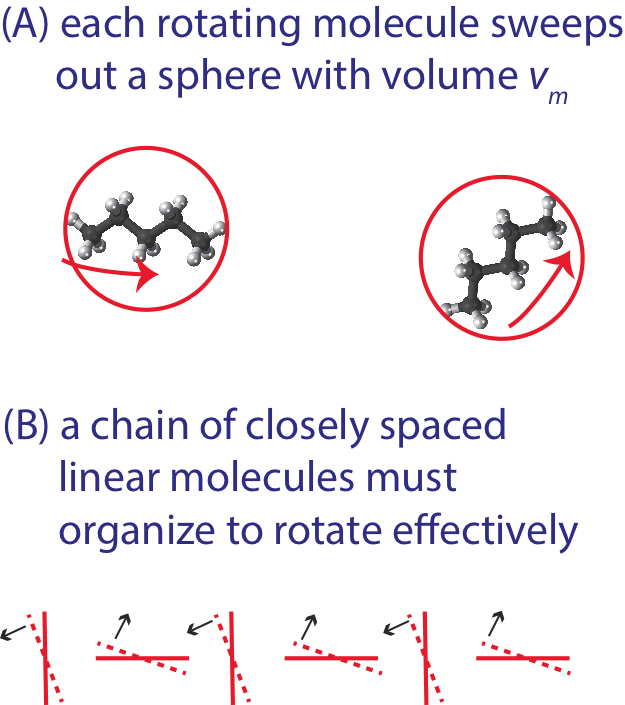}
\caption{Two possible physical pictures of repulsive clusters: (A) independently thermally rotating pentane molecules, each of length $a$, and occupying a roughly spherical volume $v_m \sim a^3$.  (B) a line of linear molecules, spaced closely enough to coordinate their thermal rotations to avoid mutual repulsion.  The length of such a chain should be approximately $R^{1/3}$.  Our results point to picture B as the better explanation for feature D.}
\label{fig:6}
\end{figure}

For the alkanes, we offer two possible mesoscopic pictures of what might constitute the ''repulsive clusters'' producing feature D.  Picture A has an alkane molecule in isolation, in a situation with relatively large molecular volume $v$; see Figure \ref{fig:6}A.  Let $c = 0.15$ nm be the approximate carbon-carbon bond length, and let $a$ be the approximate length of an alkane chain.  By virtue of its thermal rotation, the isolated alkane molecule sweeps out a spherical volume

\begin{equation} v_m \sim a^3. \label{140}\end{equation}

\noindent If there are $n$ c-c bonds in the alkane chain, then we have $a \sim n \times c$, and since $a$ scales up linearly with $Q$, we have

\begin{equation} v_m \propto Q^3, \label{150}\end{equation}

\noindent for $Q$ not too small.

\par
Picture A thus has the repulsive clusters consisting of individual, widely spaced, molecules, with their constituent atoms repelling each other by the rules of quantum mechanics.  However, if picture A is to be the correct physical interpretation for feature D, then we must have $R_{max} \sim v_m$, as shown in Fig. \ref{fig:3}d.  But below we found $R_{max} \gg v_m$ for all of the alkanes, contrary to picture A, so picture A is not consistent with our data.

\par
Picture B has the molecules more closely spaced, so that they have to rotate in tandem to avoid mutual repulsion, as shown in Fig. \ref{fig:6}B.  To organize such a gearing action requires molecular volume $v \sim v_m$, since if $v_m \gg v$ the molecules are too close and tangled to rotate effectively, and if $v_m \ll v$ the molecules are too far apart to interact significantly.  We would expect $R^{1/3}$ to give the length of a coordinated geared chain, and we might expect $R^{1/3}$ to become quite large.  Below, we found that for the alkane molecules, the conditions obtain for picture B: $v \sim v_m$ and $R_{max} \gg v_m$.  Picture B might thus have physical validity.

\par
Zwanzig has analyzed a network system of gears on a two-dimensional lattice  \cite{Zwanzig1987}.  Whether or not such a mechanism could readily reveal itself in computer simulations in fluids is unclear.

\section{Results}

In this section we present our results based on the study of the NIST/REFPROP database \cite{NIST, Lemmon2014}.  NIST/REFPROP interpolates real fluid thermodynamic data with the most accurate equations of state, representing a combination of multiparameter fitting functions and simplified fluid models.  Each fluid in the database is represented by its molar Helmholtz free energy as a function of $(T,\rho)$, yielding all of the thermodynamic quantities, including $R$.  The current edition (9.1) includes 121 pure fluids.

\subsection{Search protocol and tabulation}

For each of the $121$ pure fluids in NIST/REFPROP, we searched for Dfluids by tabulating $R$ versus $T$ in the saturated vapor phase.  Where possible, our search for positive $R$ extended from the triple point to very near the critical point, yielding $97$ Dfluids.  For each Dfluid, we examined $R$ on numerical grids abutting the saturated vapor phase.  As best as we could tell, the maximum value $R_{max}$ in feature D was always located on the saturated vapor curve.  We also found a few fluids with positive $R$ in the saturated vapor phase near the triple point (e.g., neon).  But near the triple point, the pressure and the vapor density are both typically very low, and obtaining high quality thermodynamic data can be difficult.  Hence, we include no discussion of positive $R$ near triple points.

\par
The full tabulation of Dfluids is given in Table \ref{tab:1} (in Appendix 1).  Table \ref{tab:2} (in Appendix 1) lists the 24 fluids lacking feature D.  Since the presence of feature D was found to correlate with increasing molecular complexity, Table \ref{tab:2} contains mostly simple molecules.

\subsection{Interpretation of collective results}

\par
\begin{figure}
\centering
\includegraphics[width=7cm]{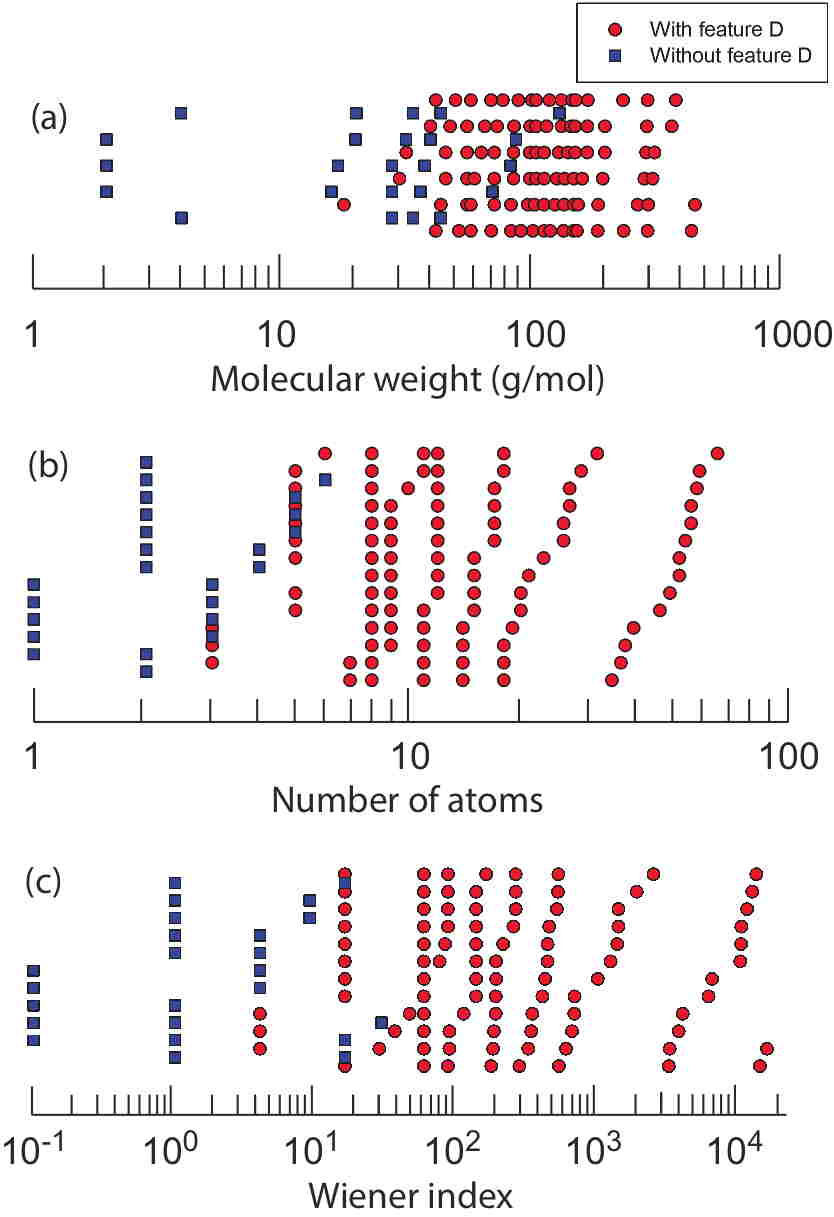}
\caption{The presence/absence of feature D, correlated with three measures of molecular complexity: (a) $MW$, (b) $Q$, and (c) $Wi$.  The vertical scales are arbitrary, and intended only to produce visual separations between nearby points.  For the five monatomic atoms ($Wi=0$) in the database, we arbitrarily set $Wi=0.1$ to display them on our logarithmic axis.}
\label{fig:Figure7}
\end{figure}

Figure \ref{fig:Figure7} indicates the presence/absence of feature D as functions of our three measures of molecular complexity.  Each measure shows a region of overlap between fluids with/without feature D.  Since we have no firm theoretical basis for picking any particular complexity measure over another, we simply preferred measures which minimize the region of overlap in Fig. \ref{fig:Figure7}.  $MW$ has 75/121 fluids in the region of overlap (a number inflated by the heavy outlier xenon), $Q$ and $Wi$ both have 23/121 overlaps.  Clearly the weight independent measures $Q$ and $Wi$ produce the clearest separation between fluids with/without feature D.  Figure \ref{fig:Figure13} in Appendix 2 has the receiver operating characteristic (ROC) curves amplifying this point.

\par
\begin{figure}
\centering
\includegraphics[width=9cm]{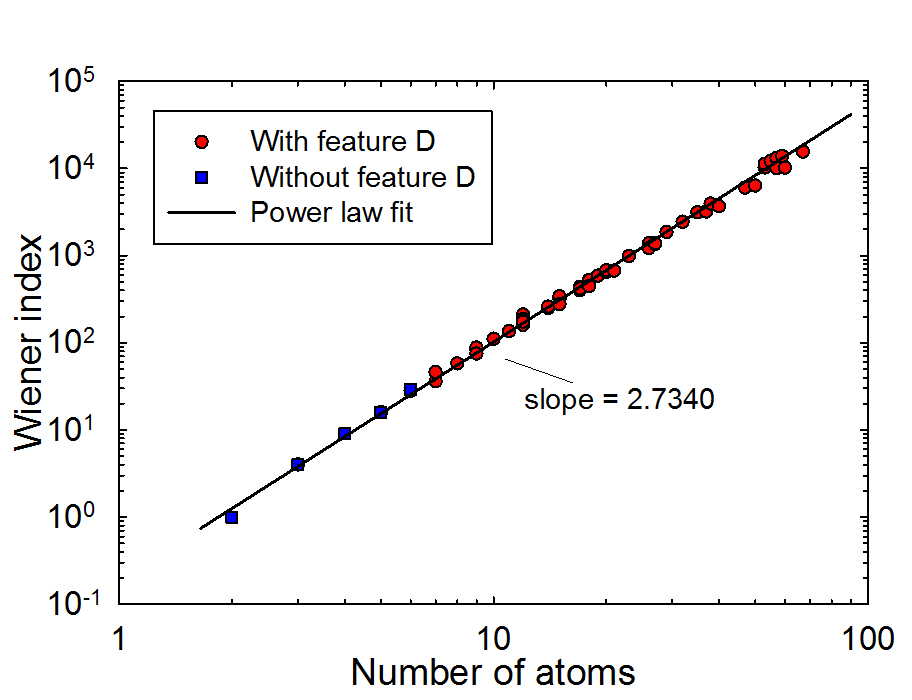}
\caption{$Wi$ versus $Q$ for all of our fluids.  The correlation between these two complexity measures is seen to be good, roughly a power law with slope 2.7340.  Fewer than the 121 data points are visible because several of the points overlap.  The five monatomic fluids (each with $Wi=0$) are not displayed.}
\label{fig:Figure8}
\end{figure}

Figure \ref{fig:Figure8} shows $Wi$ as a function of $Q$ for all of our fluids.  The correlation between $Wi$ and $Q$ is good, roughly a power law with slope $2.7340$.  As these complexity measures are equally valid, we feature the conceptually simpler $Q$.  Table 4 in Appendix 2 has Spearman ($\rho$) and Pearson ($r$) correlation coefficients calculated for all our data sets.

\par
\begin{figure}
%%%%%%%%%%%%%%%%%%%%%%%%%%%%%%
\begin{minipage}[b]{0.5\linewidth}
\includegraphics[width=2.7in]{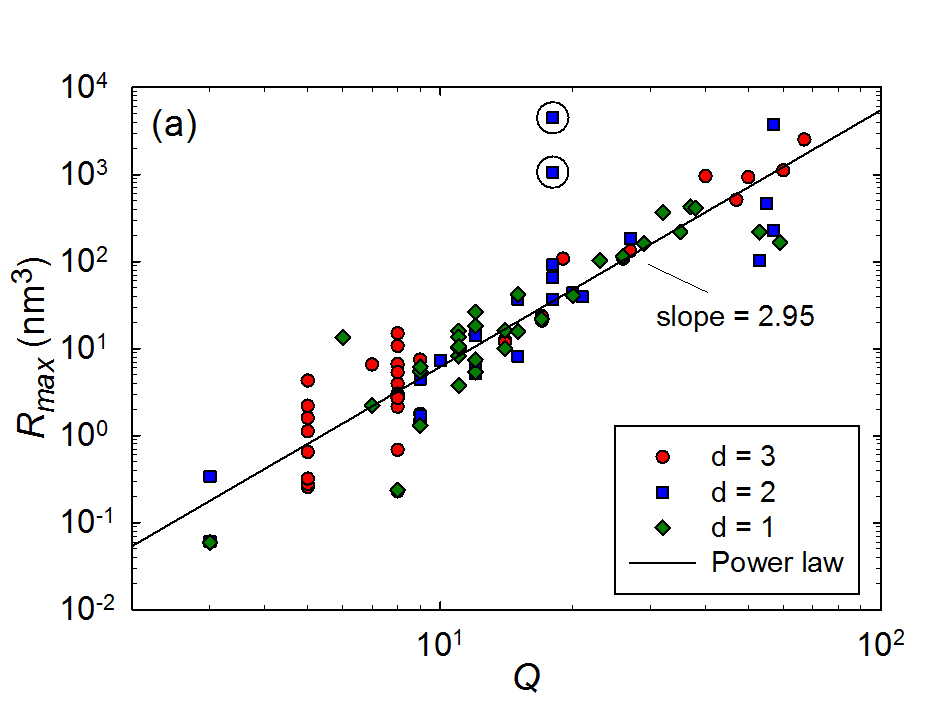}
\end{minipage}
\hspace{0.0 cm}
\begin{minipage}[b]{0.5\linewidth}
\includegraphics[width=2.7in]{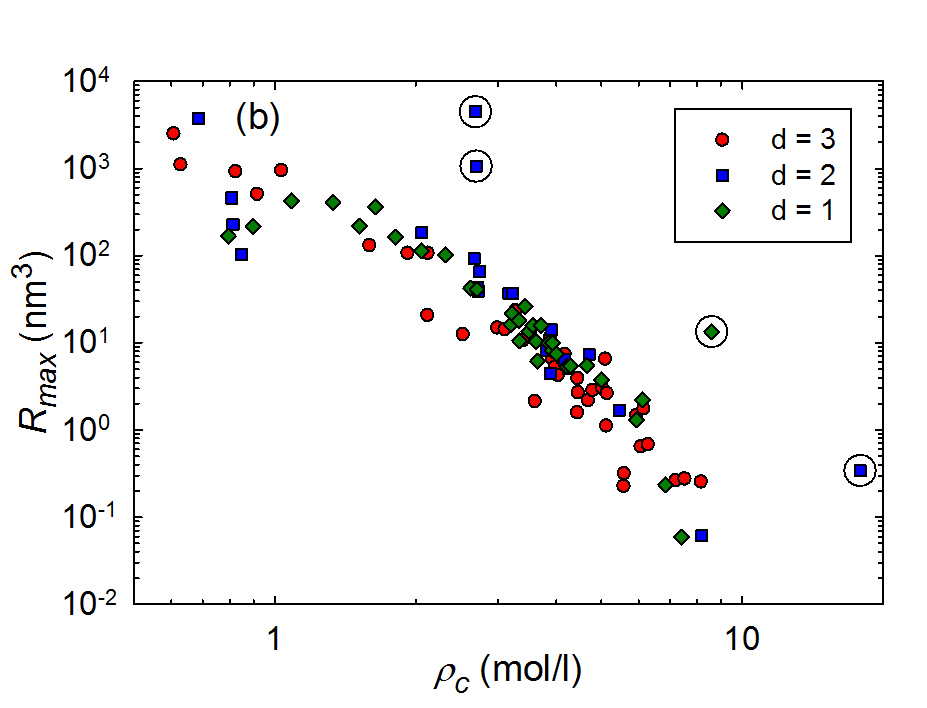}
\end{minipage}
%%%%%%%%%%%%%%%%%%%%%%%%%%%%%%%
\caption{The maximum thermodynamic curvature $R_{max}$ in feature D of each Dfluid: (a) $R_{max}$ trends upwards with increasing $Q$ roughly as a power law with slope $2.95$; and (b) $R_{max}$ trends downwards with increasing $\rho_c$.  Little visual correlation between $R_{max}$ and the molecular dimension $d$ is evident in either figure.  The circled points indicate clear outliers, identified in the text.}
\label{fig:Figure9}
\end{figure}

Figure \ref{fig:Figure9}a shows $R_{max}$ versus $Q$ for the Dfluids.  $R_{max}$ tends up with increasing $Q$, in rough accord with a power law: $R_{max} = 0.006999\,\,\mbox{nm}^3\,Q^{2.947}$.  We indicate with symbol types the molecular dimension $d$.  Visual inspection shows little correlation between $R_{max}$ and $d$, in line with expectations that $R$ measures mesoscopic properties, and not microscopic ones.  This point is amplified with statistical analysis in Appendix 2.  The two circled outlier points at $Q=18$, and well above the best-fit line, are o-xylene and p-xylene.  Figure \ref{fig:Figure9}b shows $R_{max}$ as a function of $\rho_c$ for the Dfluids.  $R_{max}$ decreases with increasing $\rho_c$, with little visual dependence on $d$.  The four circled outlier points above the best fit are o-xylene, p-xylene, methanol, and water.

\par
\begin{figure}
%%%%%%%%%%%%%%%%%%%%%%%%%%%%%%
\begin{minipage}[b]{0.5\linewidth}
\includegraphics[width=2.75in]{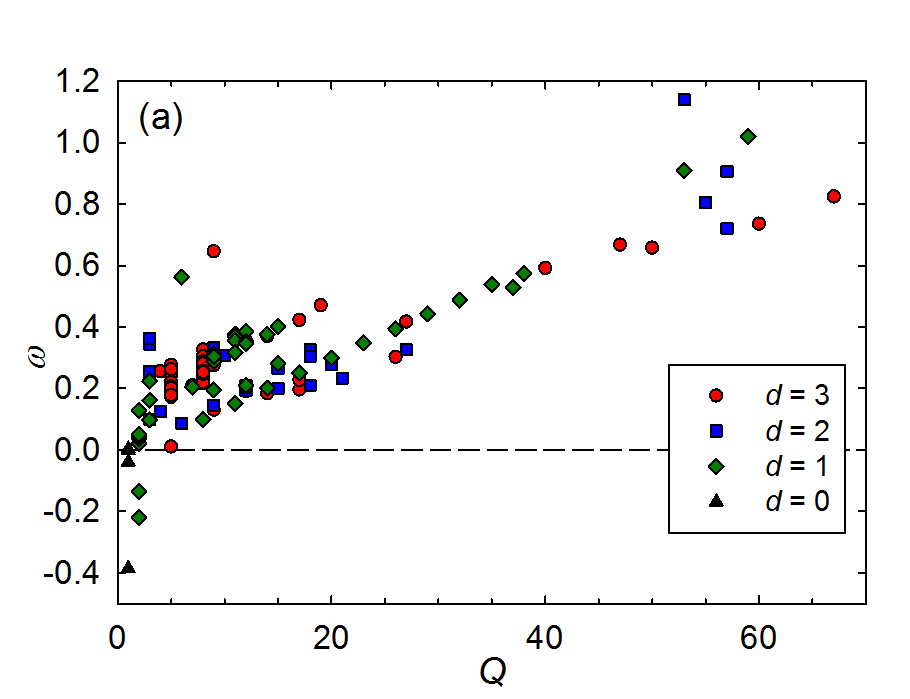}
\end{minipage}
\hspace{0.0 cm}
\begin{minipage}[b]{0.5\linewidth}
\includegraphics[width=2.8in]{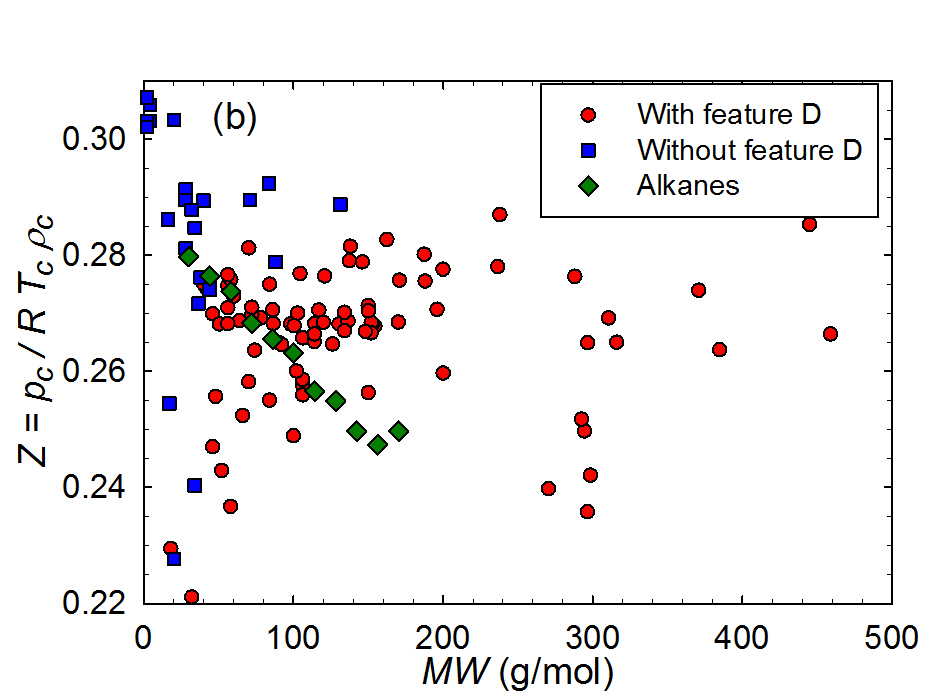}
\end{minipage}
\vspace{0.1cm}
%%%%%%%%%%%%%%%%%%%%%%%%%%%%%%%
\begin{minipage}[b]{0.5\linewidth}
\includegraphics[width=2.75in]{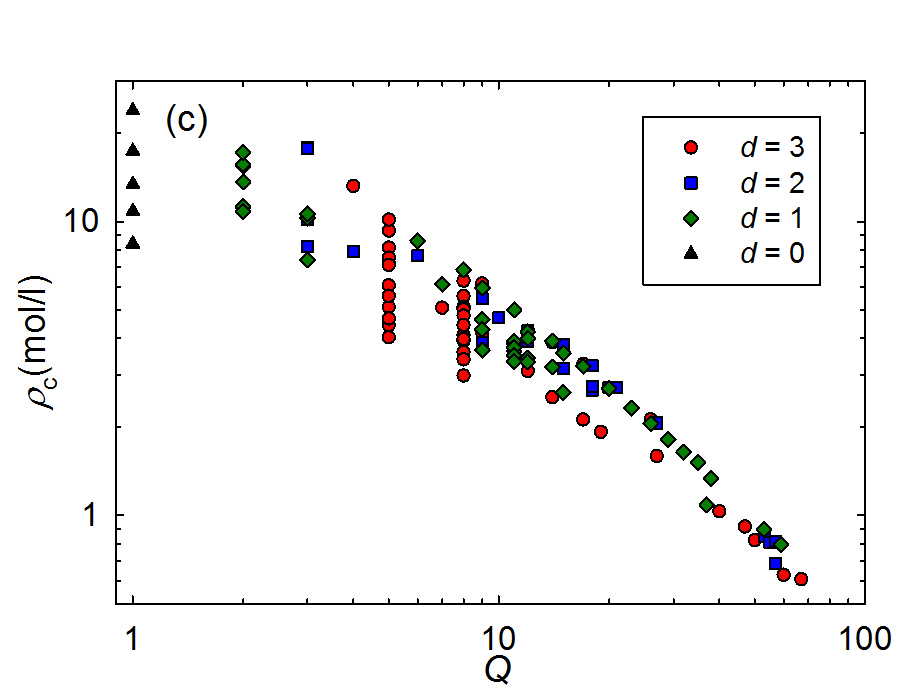}
\end{minipage}
\hspace{0.0 cm}
\begin{minipage}[b]{0.5\linewidth}
\includegraphics[width=2.83in]{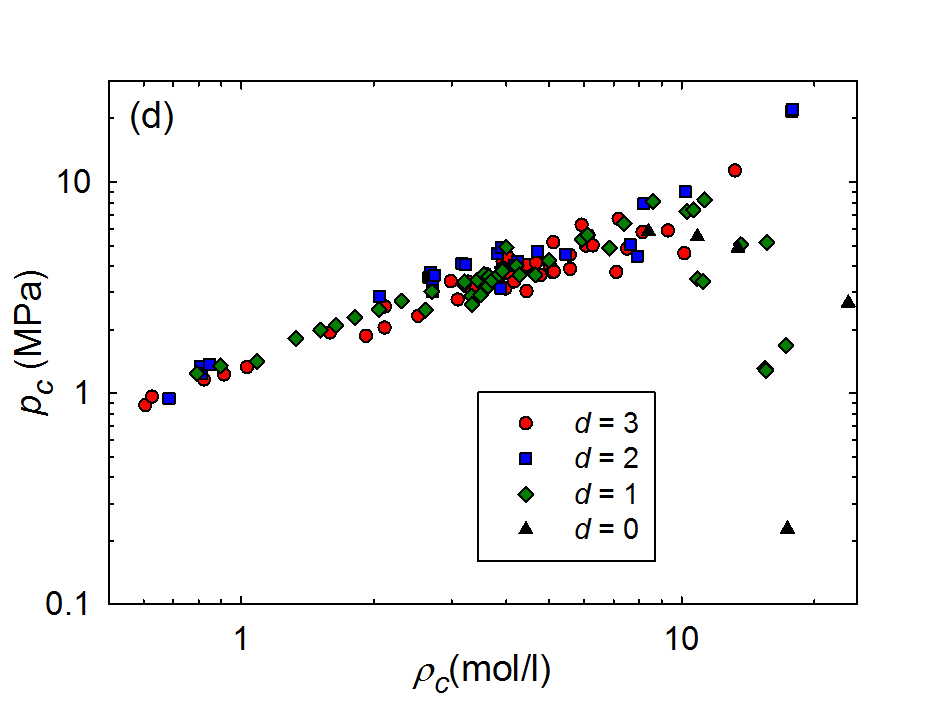}
\end{minipage}
%%%%%%%%%%%%%%%%%%%%%%%%%%%%%%%
\caption{Several thermodynamic quantities for all the $121$ fluids: (a) the acentric factor $\omega$, trending upwards with $Q$; (b) the van der Waals law of corresponding states, with values for $Z$ bunched between $0.22-0.31$; (c) the critical density $\rho_c$; and (d) the critical pressure $p_c$.  None of these figures shows clear visual dependence on $d$.}
\label{fig:Figure10}
\end{figure}

Before we discuss feature D further, we display some additional relationships in Figure \ref{fig:Figure10}.  Fig. \ref{fig:Figure10}a shows the acentric factor $\omega$ for all the 121 fluids.  Other than helium's $\omega=-0.385$, the other four noble gasses in NIST/REFPROP have $\omega$ very small.  Overall, $\omega$ clearly trends up with $Q$, so it does measure molecular complexity.  However, other than the fact that larger $Q$'s more readily allow more complicated molecular shapes, there is no particularly strong visual correlation in Fig. \ref{fig:Figure10}a between $\omega$ and $d$, perhaps surprising.

\par
We graph as well the dimensionless

\begin{equation} Z=\frac{p_c}{R\,T_c\,\rho_c} \label{160}\end{equation}

\noindent versus the molecular weight $MW$ (where $R$ is here the universal gas constant).  According to the vdW law of corresponding states, we expect a universal value $Z=3/8=0.375$ \cite{Pathria2011}.  Fig. \ref{fig:Figure10}b shows that for all of our fluids the $Z$ values fall between $0.22-0.31$, a bit lower than the vdW value.  But despite the wide variation in the critical point parameters, the combination of parameters in $Z$ have values grouping together, emphasizing the power of a simple equation of state to model a wide range of fluids.

\par
Figures \ref{fig:Figure10}c and \ref{fig:Figure10}d show correlations among $\rho_c$, $p_c$, and $Q$.  For example, the molar density $\rho_c$ decreases with increasing $Q$ because larger molecules occupy more space at the critical point.

\subsection{Alkane and perfluoroalkane results}

\begin{figure}
%%%%%%%%%%%%%%%%%%%%%%%%%%%%%%
\begin{minipage}[b]{0.5\linewidth}
\includegraphics[width=2.7in]{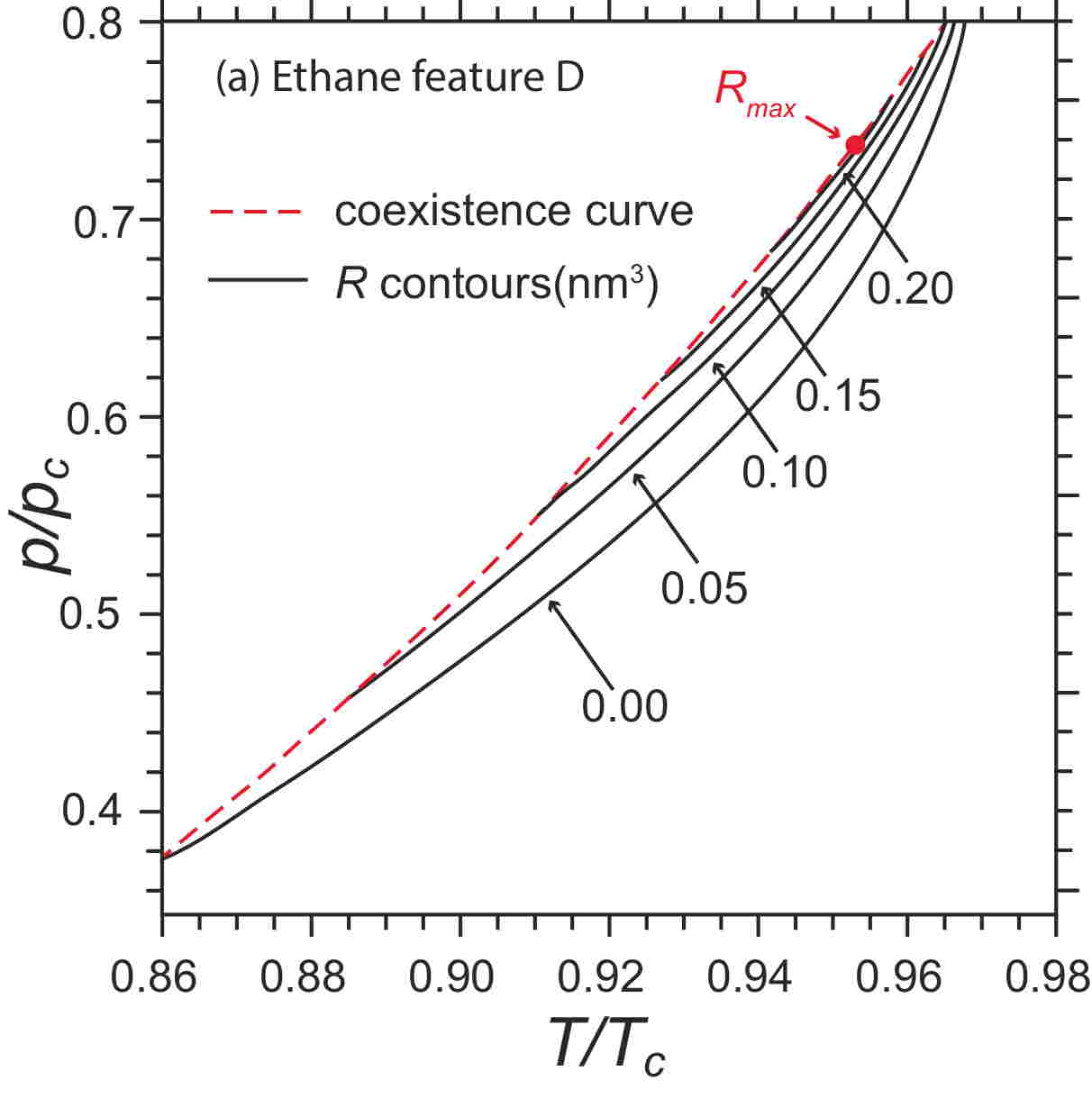}
\end{minipage}
\hspace{0.0 cm}
\begin{minipage}[b]{0.5\linewidth}
\includegraphics[width=2.7in]{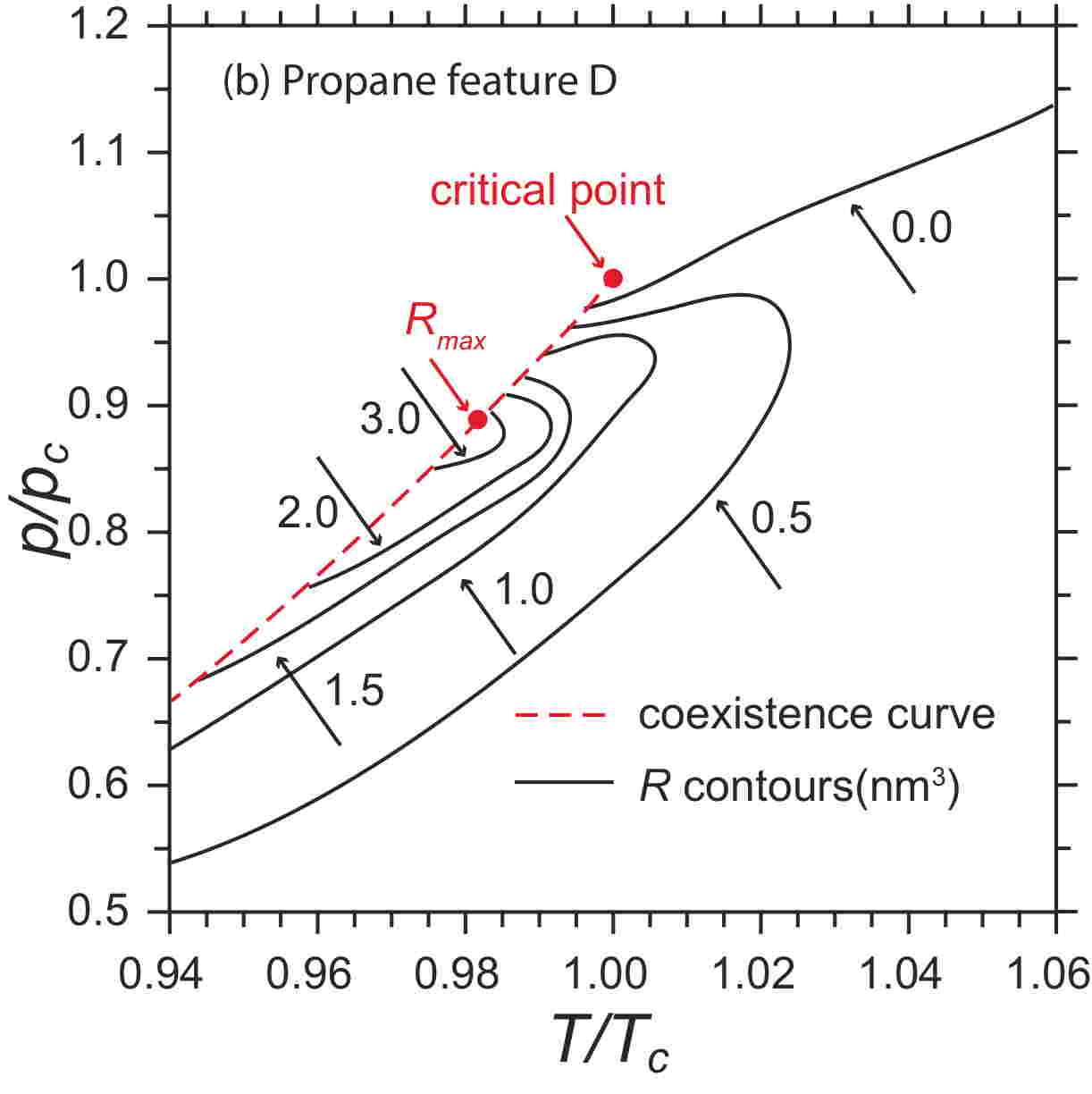}
\end{minipage}
%%%%%%%%%%%%%%%%%%%%%%%%%%%%%%%
\caption{Contour diagrams for $R$ for (a) ethane, with feature D confined to near the dense vapor phase ($R_{max}=0.236$ nm$^3$), and for (b) propane, with a more open feature D extending into the supercritical regime ($R_{max}=3.78$ nm$^3$).  The $R$ values for propane, with its larger $Q$, are larger than those for ethane.  The open feature D is characteristic of fluids with larger values of $Q$.}
\label{fig:Figure11}
\end{figure}

In this subsection we present results for the alkanes and the perfluoroalkanes, two families of linear molecular chains.  Figure \ref{fig:Figure11}a shows an $R$-diagram for ethane, with a localized feature D of a type shown in Fig. \ref{fig:4}, and also found in water \cite{Ruppeiner2015a}.  Fig. \ref{fig:Figure11}b shows an $R$-diagram for propane, with a more open feature D, encompassing portions of the supercritical region.  All of the alkanes beyond propane have features D's of this open type.

\par
Figures \ref{fig:Figure11}a and \ref{fig:Figure11}b show as well the point $\mathcal{P}_{max}$, where the maximum value $R_{max}$ was found.  Although it was difficult to be certain with our numerical grid search, to the best that we can tell it appears that $\mathcal{P}_{max}$ is always on the coexistence curve itself.  All of the values for $R_{max}$ reported in our paper are on the dense vapor phase itself.  Table \ref{tab:3} (in Appendix 1) adds information about the alkanes and the perfluoroalkanes at $\mathcal{P}_{max}$: temperature $T_{max}$, pressure $p_{max}$, and density $\rho_{max}$.

\par
\begin{figure}
%%%%%%%%%%%%%%%%%%%%%%%%%%%%%%
\begin{minipage}[b]{0.5\linewidth}
\includegraphics[width=2.75in]{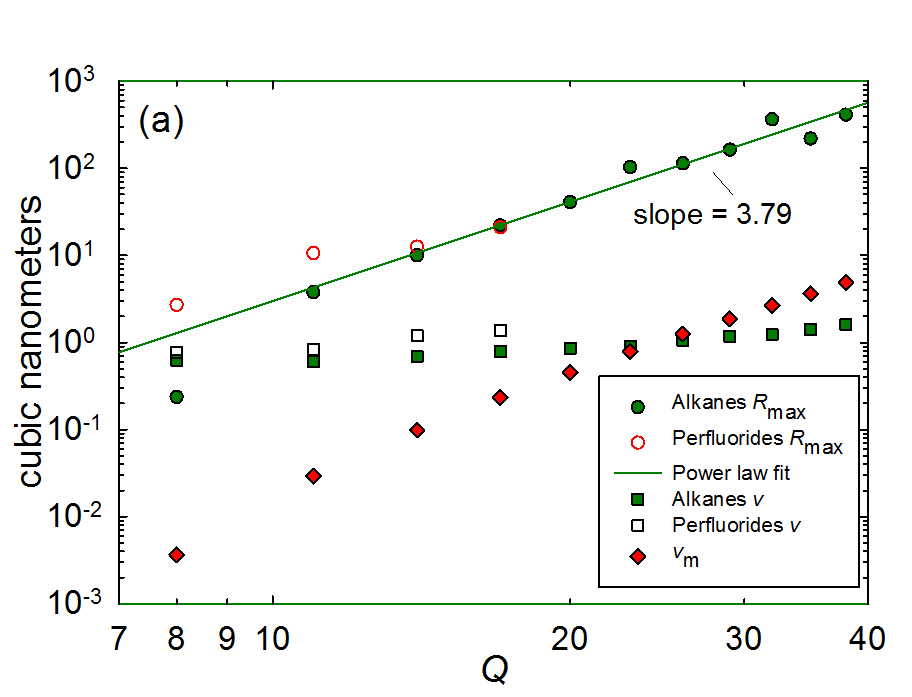}
\end{minipage}
\hspace{0.0 cm}
\begin{minipage}[b]{0.5\linewidth}
\includegraphics[width=2.8in]{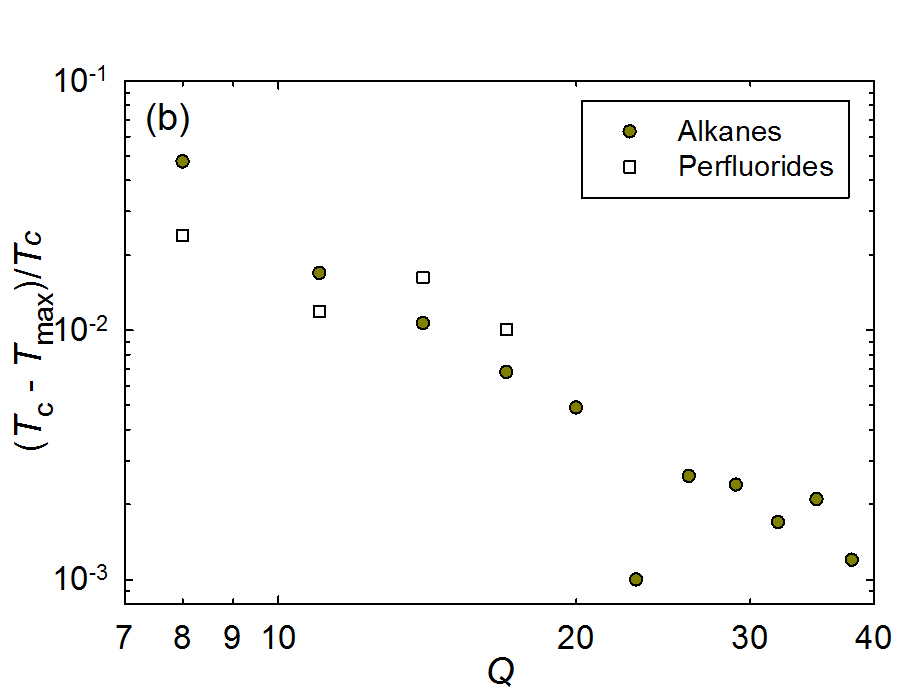}
\end{minipage}
%%%%%%%%%%%%%%%%%%%%%%%%%%%%%%%
\caption{Properties at $\mathcal{P}_{max}$: (a) $R_{max}$ for the alkanes and the perfluoroalkanes, with the molecular volume $v_{max}$, and the volume of the molecule $v_m$; (b) the reduced temperature $(T_c-T_{max})/T_c$ showing the increasing constraining of the asymptotic critical region with increasing $Q$.}
\label{fig:Figure12}
\end{figure}

Figure \ref{fig:Figure12}a shows $R_{max}$ versus $Q$ for the alkanes, as well as the molecular volume $v_{max}=\rho_{max}^{-1}$ at $\mathcal{P}_{max}$, and the volume of the molecule $v_m$ given in Eq. (\ref{140}).  Other than for ethane ($Q=8$), we clearly have $R_{max}$ much greater than both $v$ and $v_m$, speaking to organized mesoscopic structures in the dense vapor encompassing a number of molecules.  We see in addition that beyond the lighter alkanes, $v\sim v_m$, possibly enabling the collective mechanism of organized rotation we proposed in Fig. \ref {fig:6}B.  Notice also that $R_{max}$ scales up roughly the same way as $v_m$, speaking to some type of linear structure at the root of the repulsive clusters.

\subsection{Constrained critical region}

\par
Feature D is not an asymptotic critical point property, and its presence constrains the size of the asymptotic critical region.  Previously, one of us \cite{Ruppeiner2012a} proved the commensurate $R$ theorem, stating that for a given fluid in the asymptotic critical point region the values of $R$ in the coexisting liquid and vapor phases are strictly equal to each other.  Specifically:

\begin{equation}R_{l,v}=A\,|t|^{\alpha -2},\label{170}\end{equation}

\noindent with reduced temperature

\begin{equation}t=\frac{T-T_c}{T_c}\label{180},\end{equation}

\noindent heat capacity critical exponent $\alpha$, critical temperature $T_c$, and constant critical point amplitude $A$.  The subscripts ''$l,v$'' refer to values of $R$ along the liquid and vapor parts of the coexistence curve, respectively.

\par
The evaluation of $R$ in a number of systems (including the ones here) indicate that $R$ is always negative in the asymptotic critical point region.  Hence, $A$ is negative, and Eq. (\ref{170}) is inconsistent with feature D, with it's positive $R$.  Clearly, feature D cannot be in the asymptotic critical point region.

\par
These findings might resolve a long-standing question in critical phenomena: why do fluids typically have such comparatively small asymptotic critical point regions?  To quote Stanley \cite{Stanley1999}: ''But even the hand waving arguments do not explain why in some systems scaling holds for only $1-2\%$ away from the critical point and in other systems it holds for $30-40\%$ away.''  If our picture is correct, the truncated fluid asymptotic critical regions could be attributed to feature D, and the formation of its repulsive clusters.  This constraining of the critical point region by feature D is shown explicitly in Fig. \ref{fig:Figure12}b.

\par
For the ferromagnetic 1D Ising model, $R$ is uniformly negative \cite{Ruppeiner1981, Janyszek1989}.  If the ferromagnetic 3D Ising model, with a critical point in the same universality class as that of the pure fluid, can be proved likewise to have uniformly negative $R$, then it would lack any feature D, releasing the constraint on the asymptotic critical region.  We add that of the seven fluids in the asymptotic correlation length report we cited \cite{Garrabos2006} following Eq. (\ref{80}), it turns out that only SF$_6$ is a Dfluid.  The fluids picked in this correlation length study are presumably ones with fairly large asymptotic critical regions.

\section{Conclusion}

In our paper, we built on the theme of extending the pure fluid phase diagram beyond the well-known features of the sublimation curve, the melting curve, the boiling curve, and the associated triple and critical points.  Previous proposals to extend the phase diagram included the Widom line, the Fisher-Widom line, and the Frenkel line.  Our present study was made in the context of the thermodynamic curvature $R$, where it was proposed previously that curves with $R=0$ separate regions dominated by attractive/repulsive intermolecular interactions, $R<0/R>0$, respectively.

\par
Our particular focus here was on regions with positive $R$ in the dense vapor phase, ''feature D''.  Feature D was found to be present in all fluids except those consisting of the simplest molecules.  97/121 of the fluids in the NIST/REFPROP database are Dfluids.  The size $R_{max}$ of the maximum $R$ in feature D was found to increase as a power law with the number of atoms $Q$ per molecule.  Our identification of the Dfluids, and our connection of $R_{max}$ to molecular complexity are the main results of our paper.  In addition, we proposed a mesoscopic model attributing the repulsive interactions producing feature D to correlations among rotating molecules.  Such a mechanism might be searched for in computer simulations.  We also compared several measures of molecular complexity, and classified molecules according to their spatial dimension $d$.  We found that the acentric factor $\omega$ correlates with molecular complexity, but not otherwise with $d$. In addition, we displayed the strong constriction of the asymptotic fluid critical regime by feature D.

\section{Acknowledgement}  We thank Peter Mausbach for bringing to our attention the theme of nonclassical sound propagation in references \cite{Colonna2006, Colonna2009, Castier2012}, and for reference \cite{Zwanzig1987} on geared systems.

\section{Appendix 1: fluid tables}

This Appendix tabulates fluids with feature D present/absent.  Our tables list a number of fluid attributes, and we looked for correlations between all of them.  We also looked at the molecular dipole moments, but they showed no clear visual correlation with the other tabulated properties, so we did not tabulate them here.

\small

\begin{center}
\begin{longtable}{lccccccccc}
\caption{The $97$ fluids possessing feature D, along with $Q$ = number of atoms, $Wi$ = Wiener index, $MW$ = molecular weight (g/mol), $T_c$ = critical temperature (K), $p_c$ = critical pressure (MPa), $\rho_c$ = critical density (mol/l), $\omega$ = acentric factor, $d$ = molecular dimension, and $R_{max}$ = maximum thermodynamic curvature in feature D (nm$^3$).  The fluids are sorted in order of decreasing $R_{max}$.} \label{tab:1} \\

\hline
\hline

\multicolumn{1}{l}{Fluid} &
\multicolumn{1}{c}{$Q$} &
\multicolumn{1}{c}{$Wi$} &
\multicolumn{1}{c}{$MW$} &
\multicolumn{1}{c}{$T_c$} &
\multicolumn{1}{c}{$p_c$} &
\multicolumn{1}{c}{$\rho_c$} &
\multicolumn{1}{c}{$\omega$} &
\multicolumn{1}{c}{$d$} &
\multicolumn{1}{c}{$R_{max}$} \\

\hline 
\endfirsthead

\multicolumn{10}{c}
{\tablename\ \thetable{} -- continued from previous page} \\

\hline
\hline

\multicolumn{1}{l}{Fluid} &
\multicolumn{1}{c}{$Q$} &
\multicolumn{1}{c}{$Wi$} &
\multicolumn{1}{c}{$MW$} &
\multicolumn{1}{c}{$T_c$} &
\multicolumn{1}{c}{$p_c$} &
\multicolumn{1}{c}{$\rho_c$} &
\multicolumn{1}{c}{$\omega$} &
\multicolumn{1}{c}{$d$} &
\multicolumn{1}{c}{$R_{max}$} \\

\hline 
\endhead

\hline \multicolumn{10}{r}{continued on next page $\ldots$} \\
\hline
\endfoot

\hline
\hline
\endlastfoot

o-xylene & 18 & 507 & 106.2 & 630.3 & 3.738 & 2.684 & 0.312 & 2 & 4460. \\
MD3M & 57 & 10056 & 384.8 & 628.4 & 0.9450 & 0.6858 & 0.722 & 2 & 3720. \\
MD4M & 67 & 15520 & 459.0 & 653.2 & 0.8770 & 0.6061 & 0.825 & 3 & 2520. \\
D6 & 60 & 10212 & 444.9 & 645.8 & 0.961 & 0.6273 & 0.736 & 3 & 1110. \\
p-xylene & 18 & 525 & 106.2 & 616.2 & 3.532 & 2.694 & 0.324 & 2 & 1060. \\
D4 & 40 & 3688 & 296.6 & 586.5 & 1.332 & 1.031 & 0.592 & 3 & 957. \\
D5 & 50 & 6355 & 370.8 & 619.2 & 1.160 & 0.8224 & 0.658 & 3 & 928. \\
MD2M & 47 & 6012 & 310.7 & 599.4 & 1.227 & 0.9146 & 0.668 & 3 & 510. \\
methyl linoleate & 55 & 12228 & 294.5 & 799.0 & 1.341 & 0.8084 & 0.805 & 2 & 460. \\
MDM & 37 & 3188 & 236.5 & 564.1 & 1.415 & 1.085 & 0.529 & 1 & 424. \\
dodecane & 38 & 3943 & 170.3 & 658.1 & 1.817 & 1.330 & 0.574 & 1 & 411. \\
decane & 32 & 2446 & 142.3 & 617.7 & 2.103 & 1.640 & 0.488 & 1 & 365. \\
methyl oleate & 57 & 13058 & 296.5 & 782.0 & 1.246 & 0.8128 & 0.906 & 2 & 230. \\
undecane & 35 & 3136 & 156.3 & 638.8 & 1.990 & 1.515 & 0.539 & 1 & 219. \\
methyl palmitate & 53 & 10236 & 270.5 & 755.0 & 1.350 & 0.8970 & 0.91 & 1 & 217. \\
propylcyclohexane & 27 & 1356 & 126.2 & 630.8 & 2.860 & 2.060 & 0.326 & 2 & 184. \\
methyl stearate & 59 & 13836 & 298.5 & 775.0 & 1.239 & 0.7943 & 1.02 & 1 & 167. \\
nonane & 29 & 1864 & 128.3 & 594.6 & 2.281 & 1.810 & 0.443 & 1 & 163. \\
hexamethyldisiloxane & 27 & 1384 & 162.4 & 518.8 & 1.939 & 1.590 & 0.418 & 3 & 132. \\
octane & 26 & 1381 & 114.2 & 569.3 & 2.497 & 2.056 & 0.395 & 1 & 114. \\
isooctane & 26 & 1219 & 114.2 & 544.0 & 2.572 & 2.120 & 0.303 & 3 & 108. \\
novec 649 & 19 & 588 & 316.0 & 441.8 & 1.869 & 1.920 & 0.471 & 3 & 108. \\
methyl linolenate & 53 & 11250 & 292.5 & 772.0 & 1.369 & 0.8473 & 1.14 & 2 & 104. \\
heptane & 23 & 988 & 100.2 & 540.1 & 2.736 & 2.315 & 0.349 & 1 & 103. \\
m-xylene & 18 & 516 & 106.2 & 616.9 & 3.535 & 2.665 & 0.326 & 2 & 92.4 \\
ethylbenzene & 18 & 519 & 106.2 & 617.1 & 3.622 & 2.741 & 0.305 & 2 & 65.3 \\
isohexane & 20 & 649 & 86.18 & 497.7 & 3.040 & 2.715 & 0.2797 & 2 & 43.7 \\
RE347mcc & 15 & 334 & 200.1 & 437.7 & 2.476 & 2.620 & 0.403 & 1 & 42.3 \\
hexane & 20 & 676 & 86.18 & 507.8 & 3.034 & 2.706 & 0.2990 & 1 & 40.9 \\
methylcyclohexane & 21 & 672 & 98.19 & 572.2 & 3.470 & 2.720 & 0.234 & 2 & 39.2 \\
toluene & 15 & 318 & 92.14 & 591.8 & 4.126 & 3.169 & 0.2657 & 2 & 37.1 \\
cyclohexane & 18 & 447 & 84.16 & 553.6 & 4.081 & 3.224 & 0.2096 & 2 & 36.9 \\
RE245fa2 & 12 & 187 & 150.1 & 444.9 & 3.433 & 3.432 & 0.387 & 1 & 26.2 \\
neopentane & 17 & 400 & 72.15 & 433.7 & 3.196 & 3.270 & 0.1961 & 3 & 23.6 \\
isopentane & 17 & 418 & 72.15 & 460.4 & 3.378 & 3.271 & 0.2274 & 3 & 23.4 \\
pentane & 17 & 436 & 72.15 & 469.7 & 3.370 & 3.216 & 0.251 & 1 & 22.0 \\
perfluoropentane & 17 & 436 & 288.0 & 420.6 & 2.045 & 2.116 & 0.423 & 3 & 20.9 \\
RE245cb2 & 12 & 187 & 150.1 & 406.8 & 2.886 & 3.329 & 0.354 & 1 & 18.2 \\
R365mfc & 14 & 259 & 148.1 & 460.0 & 3.266 & 3.200 & 0.377 & 1 & 16.2 \\
R236ea & 11 & 136 & 152.0 & 412.4 & 3.420 & 3.716 & 0.369 & 1 & 15.9 \\
diethyl ether & 15 & 340 & 74.12 & 466.7 & 3.644 & 3.562 & 0.281 & 1 & 15.8 \\
R113 & 8 & 58 & 187.4 & 487.2 & 3.392 & 2.989 & 0.2525 & 3 & 15.0 \\
RC318 & 12 & 160 & 200.0 & 388.4 & 2.778 & 3.099 & 0.3553 & 3 & 14.4 \\
benzene & 12 & 174 & 78.11 & 562.0 & 4.907 & 3.901 & 0.211 & 2 & 14.1 \\
R227ea & 11 & 136 & 170.0 & 374.9 & 2.925 & 3.495 & 0.357 & 1 & 13.6 \\
methanol & 6 & 28 & 32.04 & 512.6 & 8.104 & 8.600 & 0.5625 & 1 & 13.4 \\
perfluorobutane & 14 & 259 & 238.0 & 386.3 & 2.323 & 2.520 & 0.371 & 3 & 12.6 \\
isobutane & 14 & 250 & 58.12 & 407.8 & 3.629 & 3.880 & 0.184 & 3 & 11.8 \\
R114 & 8 & 58 & 170.9 & 418.8 & 3.257 & 3.393 & 0.2523 & 3 & 10.7 \\
R218 & 11 & 136 & 188.0 & 345.0 & 2.640 & 3.340 & 0.3172 & 1 & 10.6 \\
R236fa & 11 & 136 & 152.0 & 398.1 & 3.200 & 3.626 & 0.3770 & 1 & 10.4 \\
R245fa & 11 & 136 & 134.1 & 427.2 & 3.651 & 3.850 & 0.3776 & 1 & 10.0 \\
butane & 14 & 259 & 58.12 & 425.1 & 3.796 & 3.923 & 0.201 & 1 & 10.0 \\
R245ca & 11 & 136 & 134.1 & 447.6 & 3.941 & 3.920 & 0.355 & 1 & 8.23 \\
cyclopentane & 15 & 275 & 70.13 & 511.7 & 4.571 & 3.820 & 0.201 & 2 & 8.15 \\
dimethyl carbonate & 12 & 211 & 90.08 & 557.0 & 4.909 & 4.000 & 0.346 & 1 & 7.47 \\
R1234yf & 9 & 86 & 114.0 & 367.9 & 3.382 & 4.170 & 0.276 & 3 & 7.45 \\
acetone & 10 & 111 & 58.08 & 508.1 & 4.700 & 4.700 & 0.3071 & 2 & 7.34 \\
R141b & 8 & 58 & 117.0 & 477.5 & 4.212 & 3.921 & 0.2195 & 3 & 6.63 \\
R124 & 8 & 58 & 136.5 & 395.4 & 3.624 & 4.103 & 0.2881 & 3 & 6.61 \\
sulfur hexafluoride & 7 & 36 & 146.1 & 318.7 & 3.755 & 5.082 & 0.21 & 3 & 6.57 \\
isobutene & 12 & 179 & 56.11 & 418.1 & 4.010 & 4.170 & 0.193 & 2 & 6.50 \\
R1233zd(E) & 9 & 86 & 130.5 & 438.8 & 3.571 & 3.650 & 0.305 & 1 & 6.19 \\
RE143a & 9 & 88 & 100.0 & 377.9 & 3.635 & 4.648 & 0.289 & 1 & 5.50 \\
R1234ze(E) & 9 & 86 & 114.0 & 382.5 & 3.635 & 4.290 & 0.313 & 1 & 5.50 \\
trans-butene & 12 & 188 & 56.11 & 428.6 & 4.027 & 4.213 & 0.21 & 1 & 5.36 \\
R115 & 8 & 58 & 154.5 & 353.1 & 3.129 & 3.980 & 0.248 & 3 & 5.34 \\
butene & 12 & 182 & 56.11 & 419.3 & 4.005 & 4.240 & 0.192 & 3 & 5.29 \\
cis-butene & 12 & 188 & 56.11 & 435.8 & 4.226 & 4.244 & 0.202 & 2 & 5.15 \\
R1216 & 9 & 86 & 150.0 & 358.9 & 3.149 & 3.889 & 0.333 & 2 & 4.46 \\
R11 & 5 & 16 & 137.4 & 471.1 & 4.408 & 4.033 & 0.1888 & 3 & 4.29 \\
R142b & 8 & 58 & 100.5 & 410.3 & 4.055 & 4.438 & 0.2321 & 3 & 3.94 \\
propane & 11 & 136 & 44.10 & 369.9 & 4.251 & 5.000 & 0.1521 & 1 & 3.78 \\
R134a & 8 & 58 & 102.0 & 374.2 & 4.059 & 5.017 & 0.3268 & 3 & 3.02 \\
R125 & 8 & 58 & 120.0 & 339.2 & 3.618 & 4.779 & 0.3052 & 3 & 2.87 \\
R116 & 8 & 58 & 138.0 & 293.0 & 3.048 & 4.444 & 0.257 & 3 & 2.70 \\
R143a & 8 & 58 & 84.04 & 345.9 & 3.761 & 5.128 & 0.2615 & 3 & 2.65 \\
propyne & 7 & 46 & 40.06 & 402.4 & 5.626 & 6.113 & 0.204 & 1 & 2.21 \\
R12 & 5 & 16 & 120.9 & 385.1 & 4.136 & 4.673 & 0.1795 & 3 & 2.19 \\
R123 & 8 & 58 & 152.9 & 456.8 & 3.662 & 3.596 & 0.2819 & 3 & 2.14 \\
cyclopropane & 9 & 75 & 42.08 & 398.3 & 5.580 & 6.143 & 0.1305 & 3 & 1.77 \\
propylene & 9 & 86 & 42.08 & 364.2 & 4.555 & 5.457 & 0.146 & 2 & 1.69 \\
trifluoroiodomethane & 5 & 16 & 195.9 & 396.4 & 3.953 & 4.431 & 0.176 & 3 & 1.59 \\
ethanol & 9 & 82 & 46.07 & 514.7 & 6.268 & 5.930 & 0.646 & 3 & 1.49 \\
dimethyl ether & 9 & 88 & 46.07 & 400.4 & 5.337 & 5.940 & 0.196 & 1 & 1.31 \\
R21 & 5 & 16 & 102.9 & 451.5 & 5.181 & 5.111 & 0.2061 & 3 & 1.12 \\
R161 & 8 & 58 & 48.06 & 375.3 & 5.010 & 6.280 & 0.216 & 3 & 0.685 \\
R22 & 5 & 16 & 86.47 & 369.3 & 4.990 & 6.058 & 0.2208 & 3 & 0.647 \\
water & 3 & 4 & 18.02 & 647.1 & 22.06 & 17.87 & 0.3443 & 2 & 0.341 \\
R13 & 5 & 16 & 104.5 & 302.0 & 3.879 & 5.580 & 0.1723 & 3 & 0.319 \\
R23 & 5 & 16 & 70.01 & 299.3 & 4.832 & 7.520 & 0.263 & 3 & 0.277 \\
R40 & 5 & 16 & 50.49 & 416.3 & 6.677 & 7.194 & 0.243 & 3 & 0.264 \\
R32 & 5 & 16 & 52.02 & 351.3 & 5.782 & 8.150 & 0.2769 & 3 & 0.256 \\
ethane & 8 & 58 & 30.07 & 305.3 & 4.872 & 6.860 & 0.0995 & 1 & 0.236 \\
R152a & 8 & 58 & 66.05 & 386.41 & 4.517 & 5.571 & 0.2752 & 3 & 0.228 \\
sulfur dioxide & 3 & 4 & 64.06 & 430.6 & 7.884 & 8.195 & 0.2557 & 2 & 0.061 \\
carbonyl sulfide & 3 & 4 & 60.08 & 378.8 & 6.370 & 7.410 & 0.0978 & 1 & 0.059 \\

\end{longtable}
\end{center}

\normalsize

\begin{center}
\begin{longtable}{lcccccccc}
\caption{The $24$ fluids not possessing feature D, along with $Q$ = number of atoms, $Wi$ = Wiener index, $MW$ = molecular weight (g/mol), $T_c$ = critical temperature (K), $p_c$ = critical pressure (MPa), $\rho_c$ = critical density (mol/l), $\omega$ = acentric factor, and $d$ = molecular dimension.  The fluids are sorted in order of descending $Q$.}  \label{tab:2} \\

\hline
\hline

\multicolumn{1}{l}{Fluid} &
\multicolumn{1}{c}{$Q$} &
\multicolumn{1}{c}{$Wi$} &
\multicolumn{1}{c}{$MW$} &
\multicolumn{1}{c}{$T_c$} &
\multicolumn{1}{c}{$p_c$} &
\multicolumn{1}{c}{$\rho_c$} &
\multicolumn{1}{c}{$\omega$} &
\multicolumn{1}{c}{$d$} \\

\endfirsthead

\multicolumn{9}{c}
{\tablename\ \thetable{} -- continued from previous page} \\

\hline
\hline

\multicolumn{1}{l}{Fluid} &
\multicolumn{1}{c}{$Q$} &
\multicolumn{1}{c}{$Wi$} &
\multicolumn{1}{c}{$MW$} &
\multicolumn{1}{c}{$T_c$} &
\multicolumn{1}{c}{$p_c$} &
\multicolumn{1}{c}{$\rho_c$} &
\multicolumn{1}{c}{$\omega$} &
\multicolumn{1}{c}{$d$} \\

\hline
\endhead

\hline
\multicolumn{9}{r}{continued on next page $\ldots$} \\
\hline
\endfoot

\hline
\hline
\endlastfoot

\hline

ethylene & 6 & 29 & 28.05  & 282.4 & 5.042 & 7.637 &  0.0866 & 2 \\
R41 & 5 & 16 & 34.03 &  317.3 & 5.897 & 9.3 & 0.2004 &  3 \\
R14 & 5 & 16 & 88.01 &  227.5 & 3.750 & 7.109 & 0.1785  & 3 \\
methane & 5 & 16 & 16.04  & 190.6 & 4.599 & 10.14 &  0.01142 & 3 \\
nitrogen trifluoride & 4  & 9 & 71.02 & 234.0 & 4.461 &  7.92 & 0.126 & 2 \\
ammonia & 4 & 9 & 17.03 &  405.4 & 11.33 & 13.21 & 0.256  & 3 \\
nitrous oxide & 3 & 4 &  44.01 & 309.5 & 7.245 & 10.27  & 0.162 & 1 \\
hydrogen sulfide & 3 & 4  & 34.08 & 373.1 & 9.000 & 10.19 &  0.1005 & 2 \\
heavy water & 3 & 4 &  20.03 & 643.9 & 21.67 & 17.78  & 0.364 & 2 \\
carbon dioxide & 3 & 4 &  44.01 & 304.1 & 7.377 & 10.62  & 0.2239 & 1 \\
parahydrogen & 2 & 1 &  2.016 & 32.94 & 1.286 & 15.54  & -0.219 & 1 \\
oxygen & 2 & 1 & 32.00 &  154.6 & 5.043 & 13.63 & 0.0222  & 1 \\
orthohydrogen & 2 & 1 &  2.016 & 33.22 & 1.311 & 15.45  & -0.218 & 1 \\
nitrogen & 2 & 1 & 28.01  & 126.2 & 3.396 & 11.18 &  0.0372 & 1 \\
hydrogen chloride & 2 & 1  & 36.46 & 324.6 & 8.263 &  11.27 & 0.1288 & 1 \\
hydrogen & 2 & 1 & 2.016  & 33.15 & 1.296 & 15.51 &  -0.219 & 1 \\
fluorine & 2 & 1 & 38.00 &  144.4 & 5.172 & 15.60 & 0.0449  & 1 \\
deuterium & 2 & 1 & 4.028  & 38.34 & 1.680 & 17.23 &  -0.136 & 1 \\
carbon monoxide & 2 & 1 &  28.01 & 132.9 & 3.494 & 10.85  & 0.05 & 1 \\
xenon & 1 & 0 & 131.3 &  289.7 & 5.842 & 8.4 & 0.0036 &  0 \\
neon & 1 & 0 & 20.18 &  44.49 & 2.679 & 23.88 &  -0.0387 & 0 \\
krypton & 1 & 0 & 83.80 &  209.5 & 5.525 & 10.85 &  -0.0009 & 0 \\
helium & 1 & 0 & 4.003 &  5.195 & 0.2276 & 17.38 &  -0.385 & 0 \\
argon & 1 & 0 & 39.95 &  150.7 & 4.863 & 13.41 &  -0.0022 & 0 \\

\end{longtable}
\end{center}

\newpage

\begin{table}[h!]
\caption{The alkanes and the perfluoroalkanes, with properties at the point $\mathcal{P}_{max}$ in feature D with maximum $R$: $Q$, $T_{max}/T_c$, $p_{max}/p_c$, $\rho_{max}$ (mol/l), and $R_{max}$ (nm$^3$).}
\label{tab:3}
\centering
\begin{tabular}{lccccc}\\
\hline
\hline
fluid & $Q$ & $T_{max}/T_c$ & $p_{max}/p_c$ & $\rho_{max}$ & $R_{max}$ 	  \\
\hline

 ethane & 8 & 0.9524 & 0.7350 & 2.642 & 0.236 \\
 propane & 11 & 0.9830 & 0.8923 & 2.748 & 3.78 \\
 butane & 14 & 0.9893 & 0.9278 & 2.385 & 10.0 \\
 pentane & 17 & 0.9932 & 0.9518 & 2.114 & 22.0 \\
 hexane & 20 & 0.9951 & 0.9676 & 1.919 & 40.9 \\
 heptane & 23 & 0.9990 & 0.9914 & 1.820 & 103. \\
 octane & 26 & 0.9974 & 0.9805 & 1.569 & 114. \\
 nonane & 29 & 0.9976 & 0.9810 & 1.412 & 163. \\
 decane & 32 & 0.9983 & 0.9855 & 1.332 & 365. \\
 undecane & 35 & 0.9979 & 0.9811 & 1.179 & 219. \\
 dodecane & 38 & 0.9988 & 0.9899 & 1.031 & 411. \\
 
\hline

R116 & 8 & 0.9760 & 0.8406 & 2.160 & 2.70 \\
R218 & 11 & 0.9881 & 0.9136 & 1.998 & 10.6 \\
perfluorobutane & 14 & 0.9837 & 0.8842 & 1.396 & 12.6 \\
perfluoropentane & 17 & 0.9899 & 0.9178 & 1.206 & 20.9 \\

\hline
\hline
\end{tabular}
\end{table}

\newpage

\section{Appendix 2: Statistical methods}

In this Appendix, we present supplementary statistical analysis for our graphs.

\par
\begin{figure}
%%%%%%%%%%%%%%%%%%%%%%%%%%%%%%
\begin{minipage}[b]{1.5\linewidth}
\includegraphics[width=1.80 in]{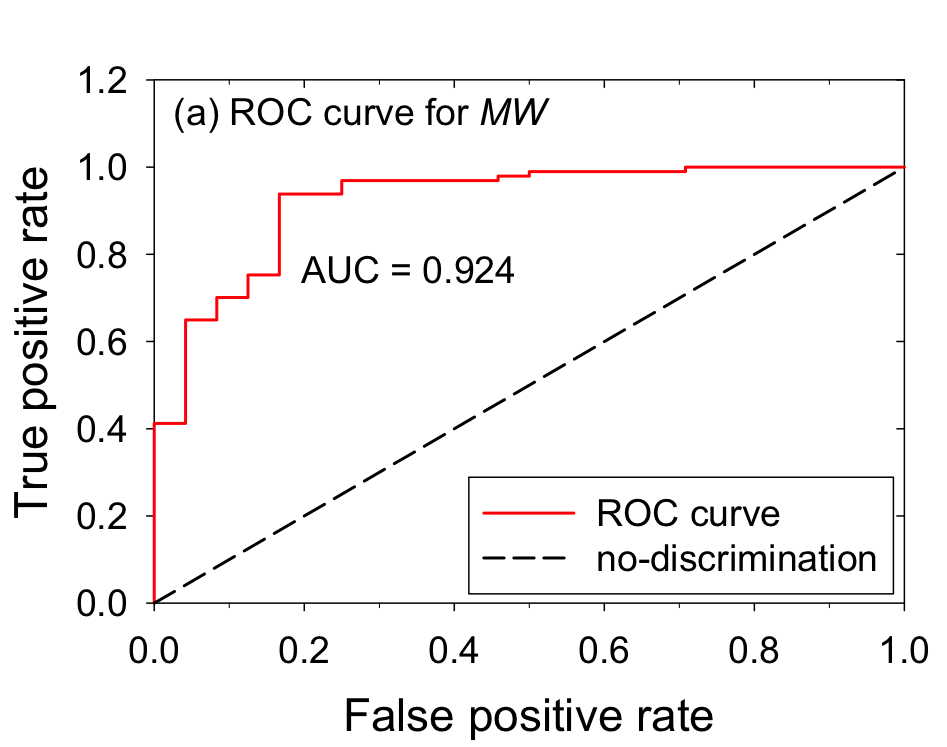}
\hspace{-0.2 cm}
\includegraphics[width=1.80 in]{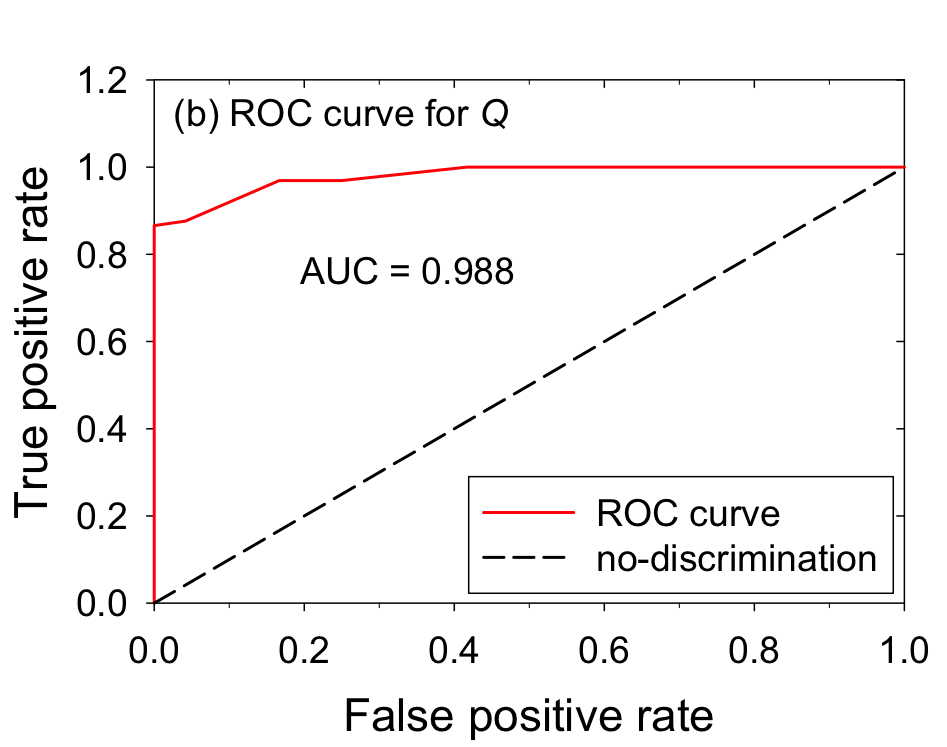}
\hspace{-0.2 cm}
\includegraphics[width=1.80 in]{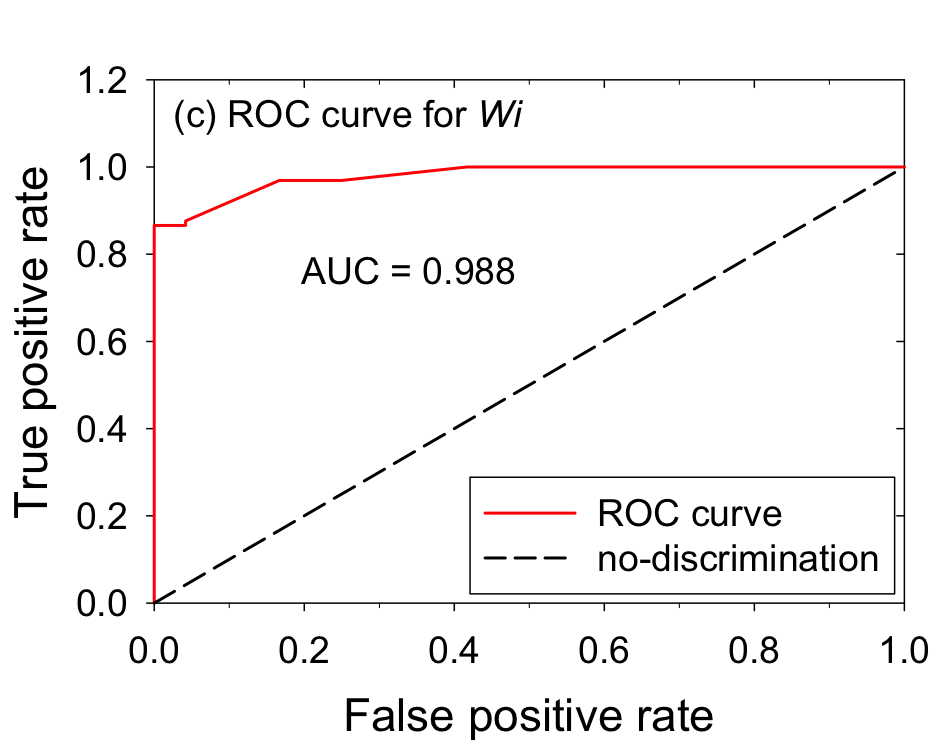}
\vspace{0.1 cm}
\end{minipage}
%%%%%%%%%%%%%%%%%%%%%%%%%%%%%%%
\caption{ROC curves for our three complexity measures for the presence/absence of feature D.  The basis for these curves are the data in Fig. \ref{fig:Figure7}.  A completely correct predictive model has area under the curve (AUC) of $1.0$.  Clearly, the weight independent measures $Q$ and $Wi$ yield the best results.}
\label{fig:Figure13}
\end{figure}

The data shown in Fig. \ref{fig:Figure7} relates the presence/absence of feature D to three possible measures of complexity.  Since we have no firm theoretical basis for preferring any one complexity measure to another, we assessed the respective classification strengths of $MW$, $Q$, and $Wi$ empirically.  We calculated the receiver operating characteristic curves (ROC) of the data in Fig. \ref{fig:Figure7}, with the results shown in Figure \ref{fig:Figure13}.

\par
Generally, ROC curves plot the true positive rate against the false positive rate at various threshold settings in the classification model.  ROC curves indicate the performances of binary classifiers.  Very accurate predictive curves follow the left-hand graph border to the point $(0,1)$, and then follow the top border of the ROC graph to the point $(1,1)$, and have area under the curve (AUC) near unity \cite{Bamber1975}.  An ROC curve closely following the 45-degree diagonal line of no-discrimination represents a completely random binary classifier, with no predictive power.

\par
Binomial logistic classification models for each measure were generated in the $R$ statistical programming language \cite{Rprogram2013}, with parameters fit by iteratively reweighted least squares regression.  The ROC curves, as well as their AUC's, were computed using the ROCR package \cite{Sing2005}.  As demonstrated in Figure \ref{fig:Figure13}, all three complexity measures demonstrate excellent classification power for feature D, with the best being $Q$ and $Wi$, as we also concluded from Fig. \ref{fig:Figure7}.

\par
Spearman ($\rho$) and Pearson ($r$) correlation coefficients, both of which fall in the range $[-1,1]$, were calculated to quantify the correlations depicted in figures \ref{fig:Figure8}, \ref{fig:Figure9}, \ref{fig:Figure10}, and \ref{fig:Figure12}, with values given in Table \ref{tab:4}.  These correlations were computed in the $R$ statistical programming language \cite{Rprogram2013}.

\begin{table}[h!]
\caption{The Spearman ($\rho$) coefficient and the Pearson ($r$) coefficient for the data point sets in our figures.}
\label{tab:4}
\centering
\begin{tabular}{lccccccc}\\
\hline
\hline
		&	F8		&	F9a		&	F9b		&	F10a		&	F10b		&	F10c		&	F10d		\\
\hline
$\rho$	&	0.999	&	0.915	&	-0.940	&	0.748	&	-0.258	&	-0.946	&	0.672	\\
$r$		&	0.911	&	0.479	&	-0.308	&	0.834	&	-0.186	&	-0.621	&	0.486	\\
\hline
\hline
		&	F12aAR	&	F12aPR	&	F12aAV	&	F12aPV	&	F12aVm	&	F12bA	&	F12bP	\\
\hline
$\rho$	&	0.991	&	1.000	&	0.991	&	1.000	&	1.000	&	-0.855	&	-0.800	\\
$r$		&	0.900	&	0.977	&	0.980	&	0.970	&	0.927	&	-0.723	&	-0.778	\\
\hline
\hline
\end{tabular}
\end{table}

\par
The data shown in Figs. \ref{fig:Figure9} and \ref{fig:Figure10} qualitatively demonstrate that the molecular dimension $d$ of a Dfluid does not strongly associate with other Dfluid features.  To quantify the strength of these relationships, the intraclass correlation (ICC) for each Dfluid attribute in our paper was calculated using the $R$ package "ICC" \cite{Wolak2012}.  The ICC is an inferential statistic that conveys the relatedness of measurements classifying naturally into groups.  An ICC value of 1 indicates that the measurements for each grouping are perfectly similar within a group \cite{Thomas1978}.  An ICC value of $0$ (or negative) indicates complete noise.  In our analysis, the groups were classified according to their $d$ value, and the measurements were the Dfluid attributes.

\par
\begin{figure}
\centering
\includegraphics[width=14cm]{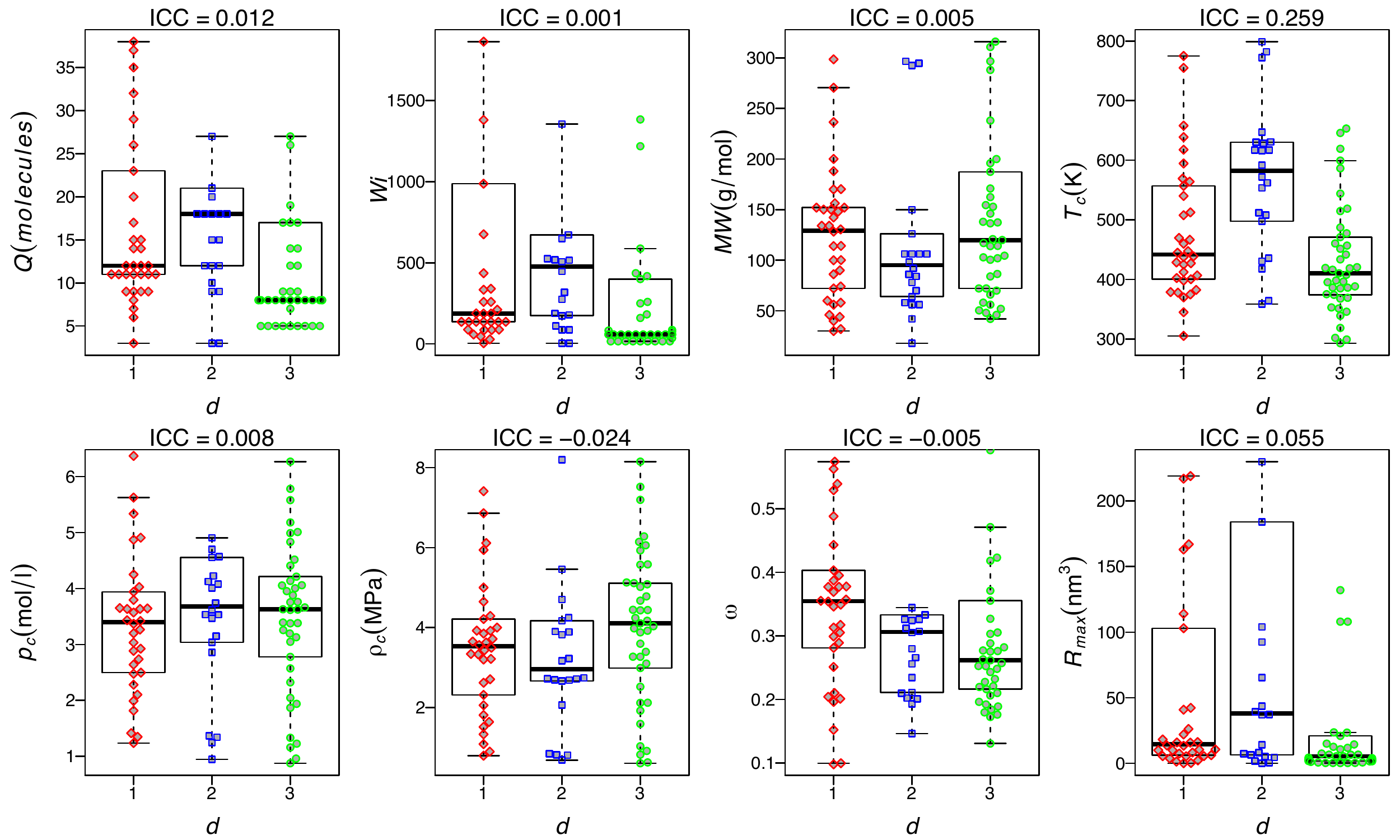}
\caption{Box plots of the calculated Dfluid attributes from Table \ref{tab:1} grouped by the molecular spatial dimension $d$. For each Dfluid attribute, the intraclass correlation (ICC) with respect to $d$ is calculated, and given above the corresponding box plot.  An ICC less than $\sim 0.4$ indicates poor similarity for the measurements within a group. Clearly, Dfluid attributes are (at best) very weakly associated with $d$.}
\label{fig:Figure14}
\end{figure}

As expected, the ICC values shown in Figure \ref{fig:Figure14} indicate that $d$ very weakly (or even negatively) associates with the other characteristic Dfluid data.  Even the measurement $T_c$ that demonstrates greatest ICC ($=0.259$) with $d$ is unconvincing.  Box plots of each Dfluid attribute grouped according to $d$, and the corresponding ICC's, are given in Fig. \ref{fig:Figure14}.

 \end{document}